\pdfoutput=1
\documentclass[conference,compsoc]{IEEEtran}

\newif\ifpublish
\publishtrue

\usepackage[T1]{fontenc}
\usepackage{amsmath,amssymb}

\usepackage{amsthm}
\usepackage{xspace}
\usepackage{graphicx}
\usepackage{cite}
\usepackage[most]{tcolorbox}
\usepackage{booktabs}
\usepackage{enumitem}
\usepackage{tikz}
\setlist[itemize]{leftmargin=*}
\setlist[enumerate]{leftmargin=*}
\usepackage{algorithm}
\usepackage[noend]{algpseudocode}
\algnewcommand{\IfThen}[2]{\State \algorithmicif\ #1\ \algorithmicthen\ #2}
\makeatletter
\AtBeginDocument{\def\theHALG@line{\thealgorithm.\arabic{ALG@line}}}
\makeatother
\usepackage[hidelinks]{hyperref}
\usepackage[nameinlink,noabbrev]{cleveref}

\crefname{section}{Section}{Sections}
\Crefname{section}{Section}{Sections}
\crefname{figure}{Figure}{Figures}
\Crefname{figure}{Figure}{Figures}
\crefname{algorithm}{Algorithm}{Algorithms}
\crefname{ALG@line}{line}{lines}
\Crefname{ALG@line}{Line}{Lines}
\Crefname{algorithm}{Algorithm}{Algorithms}
\crefname{table}{Table}{Tables}
\Crefname{table}{Table}{Tables}
\crefname{appendix}{Appendix}{Appendices}
\Crefname{appendix}{Appendix}{Appendices}
\theoremstyle{plain}
\newtheorem{theorem}{Theorem}
\newtheorem{lemma}{Lemma}
\crefname{theorem}{Theorem}{Theorems}
\Crefname{theorem}{Theorem}{Theorems}
\crefname{lemma}{Lemma}{Lemmas}
\Crefname{lemma}{Lemma}{Lemmas}

\newcommand{\sysname}{Dugong\xspace}
\newcommand{\decider}{Decider\xspace}
\newcommand{\param}[1]{\ensuremath{\mathit{#1}}\xspace}

\newcommand{\pwavelength}{\ensuremath{W}\xspace}
\newcommand{\pvoteiscert}{\param{voteIsCert}}
\newcommand{\pcommit}{\param{commit}}
\newcommand{\pskip}{\param{skip}}
\newcommand{\pcert}{\param{cert}}
\newcommand{\plink}{\param{link}}
\newcommand{\pleaders}{\param{leaders}}
\newcommand{\ppipeline}{\param{pipeline}}
\newcommand{\pasync}{\param{async}}

\newcommand{\pdag}{\ensuremath{\mathrm{DAG}}}
\newcommand{\yes}{\ensuremath{\checkmark}}
\newcommand{\no}{\ensuremath{\times}}
\newcommand{\Commit}{\ensuremath{\mathsf{Commit}}\xspace}
\newcommand{\Skip}{\ensuremath{\mathsf{Skip}}\xspace}
\newcommand{\authors}{\ensuremath{\mathrm{authors}}}
\newcommand{\para}[1]{\smallskip\noindent\textbf{#1.}~}
\newcommand{\inst}[1]{\textsuperscript{#1}}

\newcounter{claim}
\renewcommand{\theclaim}{\textbf{C\arabic{claim}}}
\newcommand{\claimitem}[1]{\item\refstepcounter{claim}\label{#1}\theclaim:}
\crefname{claim}{Claim}{Claims}
\Crefname{claim}{Claim}{Claims}
\tcbset{highlight/.style={enhanced, breakable, colback=gray!10, frame hidden, boxrule=0pt,
                  borderline west={2pt}{0pt}{black}, left=2mm, right=2mm, top=1mm, bottom=1mm}}
\newcommand{\takeaway}[2]{\begin{tcolorbox}[highlight]\textbf{Takeaway #1}: \textit{#2}\end{tcolorbox}}

\title{\sysname: Composing Uncertified DAG Consensus}
\ifpublish
    \author{
        \IEEEauthorblockN{George Danezis\inst{1,2}, Lefteris Kokoris-Kogias\inst{1}, Alberto Sonnino\inst{1,2}}
        \IEEEauthorblockA{\inst{1}Mysten Labs \quad \inst{2}University College London (UCL)}
    }
\else
    \author{}
\fi

\begin{document}
\bstctlcite{IEEEexample:BSTcontrol}
\maketitle
\pagestyle{plain}
\thispagestyle{plain}

\begin{abstract}
      There is a significant number of uncertified DAG consensus protocols spanning partial synchrony and asynchrony as well as crash, Byzantine, and hybrid fault models, yet each protocol is designed, proved, and implemented from scratch. We introduce \sysname, a single parameterized protocol that unifies uncertified DAG consensus protocols. Its core algorithm, the \decider can implement eight existing protocols and yields four new ones by just tuning its parameters. Given the property that DAGs decouple throughput from latency \sysname presents a simple trade-off between latency and resilience to network and node faults. An operator can adapt the parameters of their protocol epoch-by-epoch to adjust the resilience without needing to develop a new protocol or worry about throughput. 
 Around the \decider, \sysname layers production feature mechanisms (adaptive leader scheduling, checkpoints, responsive liveness, and crash recovery) that provably compose with each other and with the \decider's configurations. We formalize \sysname in Lean~4, proving safety and liveness under the established definitions of Byzantine atomic broadcast, and identify and repair gaps in published proofs. Finally, we implement \sysname as a reusable Rust library paired with a discrete-event network simulator, enabling direct latency comparisons across network and fault models.
\end{abstract}

\section{Introduction}
\label{sec:intro}

State machine replication (SMR) is the foundation of fault-tolerant distributed systems and blockchains~\cite{flp,dls}. Over the last decade, consensus protocols built over a directed acyclic graph (DAG) of blocks have become an established architecture for high-throughput replication (\cref{sec:related}). Among them, uncertified DAG protocols~\cite{hashgraph,blockmania,cordial,mysticeti} have been deployed in Hedera, Sui, and IOTA because of their low latency. In DAG protocols every validator proposes a block of transactions in every round and references a subset of blocks from the preceding round. This structure separates data dissemination from transaction ordering~\cite{narwhal} leading to a sepration of concerns. For the bulk data-dissemination validators batch and broadcast transactions concurrently, while for the delicate agreement protocol causal references of the DAG encode the votes. Once the DAG is constructed, validators totally order all blocks by deterministically interpreting their local view of the graph without sending additional consensus messages~\cite{dag-rider,bullshark}.

In uncertified DAG protocols~\cite{cordial,mysticeti} validators broadcast signed blocks without certifying them and encode votes and certificates in the DAG edges. Recent works have adapted uncertified DAGs to a variety of network models and fault thresholds. Under partial synchrony~\cite{dls}, Cordial Miners~\cite{cordial} and Mysticeti~\cite{mysticeti} tolerate $f$ Byzantine faults among $n \ge 3f+1$ validators and commit in three rounds; BlueBottle~\cite{bluebottle} commits in two rounds with $n \ge 5f+1$; Nemo-Nemo~\cite{nemonemo} commits in two rounds under $c$ crash faults with $n \ge 2c+1$; and Orcaella~\cite{orcaella} generalizes two-round commitment to the hybrid model with $f$ Byzantine and $c$ crash faults at $n \ge 5f+3c+1$. Under asynchrony, Cordial Miners~\cite{cordial}, Mahi-Mahi~\cite{mahimahi}, and asynchronous BlueBottle~\cite{bluebottle} replace leader timeouts with buffer rounds and a threshold common coin~\cite{cachin2005random}. Although these protocols build on the same underlying DAG structure, each work starts largely from scratch, designing custom commit rules, writing standalone proofs and prototypes, and evaluating against only one or two baselines. Starting from scratch is error-prone, as safety violations in Black Marlin~\cite{blackmarlin} and S-Minnow~\cite{minnow} show (\cref{app:attacks}). Moreover, production deployments such as Sui and IOTA~\cite{sui-lutris} require additional surrounding protocols, including leader reputation and scheduling~\cite{carousel,hammerhead}, checkpoints~\cite{sui-lutris,orcaella}, and crash recovery, whose interaction with different consensus rules has not been analyzed.

This paper takes an integrated approach to designing and analyzing uncertified DAG consensus. Instead of starting from a specific network assumption and fault model, we identify the communication patterns shared across uncertified DAG protocols and build \sysname (\cref{sec:fabric}). At the core of \sysname is a single \decider algorithm that implements a parameterized decision rule on the DAG and unifies the ordering logic of this protocol family. By setting six rule parameters, a leader schedule, and a leader election, the \decider recovers eight existing protocols across partial synchrony, asynchrony, and crash, Byzantine, and hybrid fault models~\cite{cordial,mysticeti,mahimahi,bluebottle,orcaella,nemonemo}, and yields four new protocols: Asynchronous Orcaella, Asynchronous Nemo-Nemo, Hybrid Mysticeti ($n \ge 3f+2c+1$ under partial synchrony), and Hybrid Mahi-Mahi ($n \ge 3f+2c+1$ under asynchrony).
Across all twelve configurations, the \decider captures the trade-off between resilience and latency at the same throughput: \textbf{tolerating a stronger adversary requires more rounds of communication to reach agreement}.

Around the \decider, \sysname integrates mechanisms needed in production deployments (\cref{sec:features}). These include (a) two-tier finality checkpoints~\cite{orcaella} above the consensus decisions,  (b) leader scheduling~\cite{hammerhead, barnacle} 
alongside the \decider, (c) crash recovery, and (d) responsive liveness under partial synchrony. Since every protocol parameterizes the same algorithm we can prove each composition once and compose them with one another.

We formalize \sysname and prove its safety, liveness, and composition with the mechanisms in the Lean~4 theorem prover (\cref{sec:proofs}). We use the definitions of atomic broadcast from Sailfish~\cite{sailfish} and Sui et al.~\cite{sui-duan}.
Both existing and new configurations inherit these machine-checked guarantees. We also formalize each existing protocol on its own and check that it agrees with its \decider configuration. This formalization identified gaps in published proofs, including an omitted tie-breaking rule for BlueBottle's two-round indirect commits~\cite{odontoceti,bluebottle}, which our specification fixes.

Finally, we implement \sysname in Rust as a unified library supporting all existing protocols and new configurations. We develop a deterministic discrete-event network simulator (\cref{sec:implementation}) and tune it with real inter-region latencies from AWS servers. 
Running the same implementation across all twelve configurations allows us to evaluate latency and fault tolerance on equal terms (\cref{sec:evaluation}) without the cost of large-scale cloud deployments.

\para{Contributions} We make the following contributions:
\begin{itemize}
      \item We present \sysname and its parameterized \decider, unifying uncertified DAG consensus to recover eight existing protocols and derive four new protocol configurations (\cref{sec:fabric}).
      \item We compose production mechanisms around \sysname, including adaptive leader scheduling, checkpoints, responsive liveness, and crash recovery (\cref{sec:features}).
      \item We formalize \sysname in Lean~4~\cite{leandag}, proving, under established definitions, the safety and liveness of the \decider configurations and their composition with the feature mechanisms, while identifying and fixing gaps in published proofs (\cref{sec:proofs}).
      \item We implement \sysname in Rust~\cite{dugong-code} and evaluate all twelve configurations in a discrete-event simulator, quantifying the latency and fault-tolerance trade-offs of uncertified DAG consensus on equal terms (\cref{sec:implementation,sec:evaluation}).
\end{itemize}

\section{Background and DAG-Based Consensus}
\label{sec:background}

\para{DAG construction and local interpretation}
DAG-based consensus protocols implement atomic broadcast by decoupling transaction dissemination from ordering~\cite{narwhal}. A committee of $n$ validators (or $n$ units of stake) constructs a round-based directed acyclic graph of blocks. In each round $r$, every correct validator proposes at most one signed block containing a batch of transactions and cryptographic hash references to parent blocks from round $r-1$. A non-genesis block is syntactically valid if it references at least a quorum of $q$ blocks from round $r-1$ authored by distinct validators, includes its own author's round-$(r-1)$ block, and references at most one round-$(r-1)$ block per author. Besides its parents, every block carries \emph{weak references}: a correct validator references every earlier-round block it holds that its new block would not otherwise reach. Weak references count toward no quorum, and they ensure that every correct block eventually enters the history of every later correct block, even one that no later block takes as a parent~\cite{dag-rider,sailfish}.

A block $b'$ is in the \emph{causal history} of $b$ if $b'$ is reachable from $b$ by following parent references. Each validator maintains a local \emph{view} of the DAG, which is a causally closed subset of valid blocks. A validator adds a block to its view only after downloading and validating the block's entire causal history with the help of a block synchronizer. Once blocks are in the local view, ordering requires no extra communication. Each validator applies a deterministic decision rule to its view to commit a sequence of \emph{slots} (see below). Committing a leader block commits its entire causal history, which validators flatten in a deterministic order to produce the final transaction log~\cite{dag-rider}.

\para{Uncertified DAGs}
In uncertified DAG protocols, validators broadcast signed blocks directly, in a single message delay per round. Because blocks are not certified prior to dissemination, a Byzantine validator can \emph{equivocate} by sending different blocks for the same round to different validators. Uncertified DAG protocols decide \emph{slots} and tolerate equivocation by encoding votes, blames, and certificates implicitly in the parent references of subsequent rounds:
\begin{itemize}
      \item \textbf{Decision slots:} A slot $(r, \ell)$ is a round $r$ and a leader index $\ell$, whose leader validator is set by a schedule or, under asynchrony, a common coin. Its \emph{leader blocks} are the leader's round-$r$ blocks, several if it equivocates. Over a \emph{wave} of rounds, from proposal through vote to decision, each slot is committed (one leader block), skipped, or left undecided. Committed slots, in round order, determine the delivered sequence.
      \item \textbf{Votes and blames:} A block $b$ at a round above a leader slot $(r, \ell)$ \emph{votes} for a candidate block $L$ of that slot if $L$ is the first block for $(r, \ell)$ encountered during a deterministic depth-first traversal of $b$'s parent references. Because the traversal order is fixed and a valid block references at most one parent per author, a correct validator votes for at most one block per slot even when the leader equivocates. If $b$ does not reach any block for $(r, \ell)$, $b$ \emph{blames} the slot.
      \item \textbf{Certificates:} In three-round protocols such as Mysticeti~\cite{mysticeti}, a block $C$ two rounds above the leader's proposal \emph{certifies} $L$ if $C$'s parents include votes for $L$ from at least a certificate threshold of distinct authors. In two-round protocols such as BlueBottle~\cite{bluebottle}, Orcaella~\cite{orcaella}, and Nemo-Nemo~\cite{nemonemo}, votes act directly as certificates without a separate certification round.
\end{itemize}

\para{Network models, fault models, and trade-offs}
Uncertified DAG protocols operate under different failure and timing assumptions, trading fault tolerance against commit latency. In the hybrid fault model~\cite{orcaella}, an adversary can corrupt up to $f$ \emph{Byzantine} validators, which may equivocate or deviate arbitrarily from the protocol, and up to $c$ \emph{crashed} validators, which follow the protocol until they permanently stop sending messages. The remaining $n - f - c$ validators are \emph{correct}. Setting $c = 0$ yields the standard Byzantine fault model, while setting $f = 0$ yields the crash fault-tolerant (CFT) model. All quorum thresholds, except the skip threshold, are bounded by $n - f - c$ so that the system can make progress even when all $f + c$ faulty validators are silent. As a special case, the skip threshold may be $n$, indicating no direct skip, leaving undecided slots to be decided by the indirect rule.

For network timing, protocols assume either partial synchrony~\cite{dls} or full asynchrony~\cite{flp}. Under partial synchrony, message delays between correct validators are bounded by $\Delta$ after an unknown Global Stabilization Time (GST). Partially synchronous protocols use a predefined leader schedule and wait up to a timeout for each round's leader blocks before advancing to the next round, ensuring deterministic liveness after GST. The schedule may also adapt to the committed history~\cite{hammerhead,barnacle}, and a responsive validator advances as soon as it holds the blocks the rule counts (\cref{sec:features}). Under asynchrony, message delays are unbounded. To circumvent the FLP impossibility result~\cite{flp}, validators in asynchronous protocols advance to the next round as soon as they receive a quorum of blocks from the current round, without waiting for specific leaders, and elect each slot's leader retroactively using a threshold common coin~\cite{cachin2005random} revealed in the final round of the wave~\cite{cordial,mahimahi,bluebottle}.

\para{Problem definition}
Each validator \emph{delivers} a sequence of blocks. A protocol implements \emph{atomic broadcast}~\cite{dag-rider,sailfish,sui-duan} against up to $f$ Byzantine and $c$ crashed validators when it satisfies four properties. \emph{Agreement}: if a correct validator delivers a block $b$, every correct validator eventually delivers $b$. \emph{Integrity}: every correct validator delivers each block at most once, and delivers a block of a correct validator only if that validator built it. \emph{Validity}: every block built by a correct validator is eventually delivered by every correct validator. \emph{Total order}: if two correct validators deliver blocks $b$ and $b'$, they deliver them in the same order. It is \emph{Byzantine} atomic broadcast when $f > 0$, and crash-tolerant when $f = 0$. We also prove a stronger safety property, \emph{prefix consistency}: the delivered sequences of any two correct validators are prefixes of one another. Prefix consistency, Total order, and Integrity hold in every execution, whatever the message delays, before GST and under asynchrony alike. Agreement and Validity are liveness properties. Agreement needs only that messages between correct validators are eventually delivered. Validity holds after GST under partial synchrony, and with probability 1 under asynchrony against an adversary that controls message delays but cannot predict the common coin.

The fault and network assumptions determine the shortest wave length, namely the number of rounds needed for a direct commit. Under partial synchrony, tolerating $n \ge 3f+1$ requires three rounds per wave (propose, vote, and certify) so that any two certificate quorums intersect in at least one correct validator, and extending this three-round rule to hybrid faults requires $n \ge 3f+2c+1$.
Committing in two rounds eliminates the certificate round and therefore requires a larger committee to prevent an equivocating leader from splitting the vote quorum: $n \ge 5f+1$ under Byzantine faults~\cite{bluebottle} and $n \ge 5f+3c+1$ under hybrid faults~\cite{orcaella}. Under crash faults, no leader equivocates, and Nemo-Nemo~\cite{nemonemo} commits in two rounds with $n \ge 2c+1$ by giving up direct skips (\cref{sec:fabric:new}). Adapting any of these protocols to asynchrony preserves its committee size and quorum thresholds while inserting one or two \emph{buffer rounds} between the propose and vote rounds, so that proposals spread before the common coin reveals the leader (\cref{sec:fabric:new}).

\para{Existing uncertified DAG protocols and common message patterns}
While Cordial Miners~\cite{cordial}, Mysticeti~\cite{mysticeti}, Mahi-Mahi~\cite{mahimahi}, BlueBottle~\cite{bluebottle}, Orcaella~\cite{orcaella}, and Nemo-Nemo~\cite{nemonemo} target different network and fault assumptions, their commit rules all check the same two graph patterns over each wave:
\begin{itemize}
      \item \textbf{Direct decision rule:} At the decision round of a slot's wave, a validator examines its local view. If at least a commit threshold of distinct authors in the decision round certify a candidate block $L$, the validator directly commits $L$. Conversely, if at least a skip threshold of distinct authors in the vote round blame the slot, the validator directly skips the slot.
      \item \textbf{Indirect decision rule:} If neither threshold is met, the slot remains undecided at the end of its wave. The validator then decides the slot indirectly from the earliest subsequent wave whose leader block $A$ (the \emph{anchor}) is committed: if $A$'s causal history contains at least a link threshold of decision-round certificates for a candidate $L$ (breaking ties by the least block identifier if multiple equivocating blocks qualify), the validator commits $L$, and otherwise it skips the slot.
\end{itemize}
Across all of these protocols, safety relies on three quorum-intersection properties between the direct and indirect rules: (i) a slot that is directly skipped in one view cannot have enough certificates in an anchor's causal history to be indirectly committed in another view; (ii) a directly committed block cannot be outvoted at an anchor by a conflicting block from the same slot; and (iii) a directly committed block is guaranteed to have enough certificates in the causal history of every subsequent valid anchor to be indirectly committed by any validator.
In \cref{sec:fabric}, we capture this common decision logic in a single parameterized \decider.

\section{A Parameterized \decider}
\label{sec:fabric}

\sysname separates the parts that uncertified DAG protocols share from the parts that vary across configurations. Every configuration uses the same DAG construction and the same direct and indirect decision rules, while varying the wave shape, thresholds, parent quorum, leader schedule, and election mode. This section describes the common part, as a single \decider algorithm that takes a DAG and returns, for each slot, whether it is committed, skipped, or undecided. Unlike classical consensus protocols, the \decider is not event-driven: it is a deterministic function from a local view of the DAG to decisions, and leaves messages, timers, and synchronization to the DAG layer (\cref{sec:background}).

Every configuration of \sysname runs on the DAG layer of \cref{sec:background}. We call a block's reference to its author's previous block its \emph{self-parent}, and the number of previous-round blocks every block references, and every validator waits for before advancing, the \emph{parent quorum} $q$ (\cref{sec:fabric:params}).

\begin{figure}[t]
      \centering
      \includegraphics[width=\columnwidth]{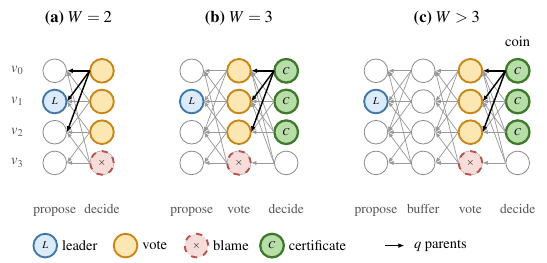}
      \caption{Wave shapes for $W=2$, $W=3$, and $W>3$ ($q = 3$).}
      \label{fig:waves}
\end{figure}

As the DAG grows, the \decider decides the slots of \cref{sec:background}, whose leaders the schedule assigns (\cref{sec:fabric:params}). Each slot spans a wave of \pwavelength consecutive rounds from its \emph{proposal round} to its \emph{decision round} (\cref{fig:waves}). Validators cast votes in the \emph{vote round}, which equals the decision round when votes act as certificates and precedes it by one round otherwise. Any rounds between the proposal round and the vote round are \emph{buffer rounds}.

\para{Challenges}
Unifying these rules into a single \decider requires resolving two issues.
First, direct and indirect decisions must agree. Protocols committing in three or more rounds~\cite{cordial,mysticeti} have unique certificates, thus a single certificate in the anchor's history suffices for safety. Without certificates, however, conflicting blocks can collect votes, and deciding indirectly via the same rule is unsafe. This is not just of theoretical difficulty as both Black Marlin~\cite{blackmarlin} and S-Minnow~\cite{minnow} suffered this issue(\cref{app:attacks}). The \decider instead treats votes as certificates and sets \plink low enough for every anchor to reach after a direct commit, yet too high for a conflicting block to clear it.
Second, existing commit rules diverge once buffer rounds or equivocating blocks appear. Counting blames right after the proposal round misses votes cast through buffer blocks. Testing skips per candidate block is unsafe: a validator that first tests an equivocating block with no votes skips the whole slot, while another validator commits the leader's other block. The \decider counts slot-level blames in the vote round and separates the parent quorum $q$ from the safety thresholds.

\subsection{The \decider}
\label{sec:fabric:decider}

\begin{figure}[t]
      \centering
      \includegraphics[width=\columnwidth]{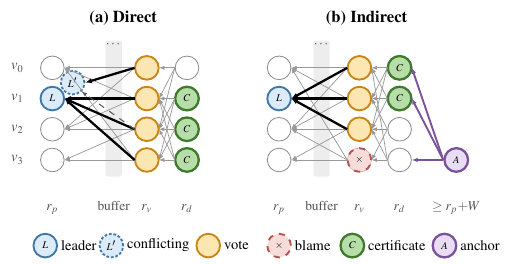}
      \caption{Direct (a) and indirect (b) rules for any wave length. In (a), the leader equivocates.}
      \label{fig:rules}
\end{figure}

\begin{algorithm}[t]
      \caption{The \decider (parameters in \cref{sec:fabric:params}).}
      \label{alg:decider}
      \footnotesize
      \begin{algorithmic}[1]
            \Statex \textbf{Notation}
            \Statex \hspace{1em}$\pdag[r]$: round-$r$ blocks in the local view
            \Statex \hspace{1em}$\authors\, X$: distinct authors of the blocks in $X$
            \Statex \hspace{1em}Decisions: $\Commit(L)$, \Skip, or $\bot$ (undecided)
            \Statex \textbf{Slots}
            \Statex \hspace{1em}\textsc{Slots} lists slots newest first, by round then leader index:
            \Statex \hspace{2em}every round if \ppipeline, else every \pwavelength-th round
            \Statex \hspace{1em}$r_p(s) = s.\mathit{round}$ \Comment{proposal round}
            \Statex \hspace{1em}$r_d(s) = r_p(s) + \pwavelength - 1$ \Comment{decision round}
            \Statex \hspace{1em}$r_v(s) = r_d(s)$ if \pvoteiscert, else $r_d(s) - 1$ \Comment{vote round}
            \Statex \hspace{1em}$L$ ranges over $s.\mathit{leader}$'s blocks in round $r_p(s)$ \Comment{leader blocks}
            \Statex \textbf{Voting}
            \Statex \hspace{1em}$\Call{Vote}{b, s}$: the block of slot $s$ that $b$'s history reaches first
            \Statex \hspace{2em}by depth-first search, or $\bot$ if none ($b$ blames $s$)
            \Statex \hspace{1em}$C$ certifies $L$: $\Call{Vote}{C, s} = L$ if \pvoteiscert, else $C$'s parents
            \Statex \hspace{2em}include votes for $L$ from at least \pcert distinct authors
            \Statex \textbf{Counting}
            \Statex \hspace{1em}$\Call{Blamers}{s} = \authors\{b \in \pdag[r_v(s)] : \Call{Vote}{b, s} = \bot\}$
            \Statex \hspace{1em}$\Call{Certs}{s, L, H} = \authors\{C \in H : C \text{ certifies } L\}$
            \Statex
            \Procedure{TryDecide}{$r_{\mathit{last}}, r_{\mathit{high}}$}
            \State $S \gets [\,]$ \Comment{decisions, oldest first}
            \For{$s \in \Call{Slots}{r_{\mathit{high}} \textbf{ down to } r_{\mathit{last}}}$} \label{line:loop}
            \State $d \gets \Call{TryDirect}{s}$ \label{line:direct}
            \IfThen{$d = \bot$}{$d \gets \Call{TryIndirect}{s, S}$} \label{line:indirect}
            \State $S \gets d \parallel S$
            \EndFor
            \State \Return longest decided prefix of $S$ \label{line:prefix}
            \EndProcedure
            \Statex
            \Procedure{TryDirect}{$s$}
            \If{$|\Call{Blamers}{s}| \ge \pskip$} \Comment{direct skip} \label{line:skip}
            \State \Return \Skip
            \EndIf
            \State \Return $\Commit(L)$ for the least $L$ with \Comment{direct commit} \label{line:commit}
            \Statex \hspace{\algorithmicindent}\hspace{1.2em}$|\Call{Certs}{s, L, \pdag[r_d(s)]}| \ge \pcommit$, else $\bot$
            \EndProcedure
            \Statex
            \Procedure{TryIndirect}{$s, S$}
            \State $a \gets$ first non-\Skip decision in $S$ at round $\ge r_p(s) + \pwavelength$ \label{line:anchor}
            \IfThen{there is none, or $a = \bot$}{\Return $\bot$} \Comment{no anchor yet}
            \State \textbf{let} $\Commit(A) = a$ \Comment{anchor $A$}
            \State \Return $\Commit(L)$ for the least $L$ with \Comment{indirect rule} \label{line:link}
            \Statex \hspace{\algorithmicindent}\hspace{1.2em}$|\Call{Certs}{s, L, \pdag[r_d(s)] \cap \mathrm{history}(A)}| \ge \plink$, else \Skip
            \EndProcedure
      \end{algorithmic}
\end{algorithm}

The \decider implements the direct and indirect rules of \cref{sec:background}. \Cref{alg:decider} gives the definitions it uses, the commit loop, and the two rules. \Cref{fig:rules} illustrates both rules for an arbitrary wave length.

\para{Commit loop}
\textsc{TryDecide} runs whenever the local view grows. It visits the slots the schedule defines (\textsc{Slots}), newest first, from the highest round $r_{\mathit{high}}$ down to $r_{\mathit{last}}$, the round of the last decision the caller consumed (\cref{line:loop}), and decides each slot by the direct rule first and the indirect rule second (\cref{line:direct,line:indirect}). The decisions accumulate in a list $S$, oldest slot first, so that every slot above the current one has been examined when the indirect rule looks for an anchor. The loop returns the longest prefix of $S$ in which every slot is decided (\cref{line:prefix}), and the caller commits the sub-DAGs of its committed leaders in order. An undecided slot therefore blocks delivery of every slot above it, so an equivocating leader delays subsequent waves in a pipelined protocol (\cref{sec:eval:byzantine}). Restarting from the last consumed round is idempotent because revisited slots keep their decisions, and the caller drops them (\cref{app:security}).

\para{Votes and blames}
A block $b$ above a slot votes for one block of that slot in its causal history, chosen by a deterministic vote rule (\textsc{Vote}). If its history holds no block of the slot, $b$ blames the slot. Without equivocation the history holds at most one block of the slot, and every rule picks it.
When the leader equivocates, $b$'s history may contain several of the leader's blocks, and \textsc{Vote} picks one, so $b$ votes for at most one block per slot. Our proofs hold for any deterministic \textsc{Vote}, and \sysname uses depth-first traversal in every configuration.

\para{Direct rule}
\textsc{TryDirect} skips the slot on \pskip blames in the vote round (\cref{line:skip}), and otherwise commits a leader block $L$ on \pcommit certificates in the decision round (\cref{line:commit}). A certificate is the vote itself when \pvoteiscert holds, and a block referencing \pcert votes for $L$ otherwise ($C$ certifies $L$ in \cref{alg:decider}). The thresholds ensure that at most one of an equivocating leader's conflicting blocks reaches \pcommit, so taking the least identifier only makes the choice definite. A slot that reaches neither threshold stays undecided ($\bot$).

\para{Indirect rule}
\textsc{TryIndirect} resolves a slot that the direct rule leaves undecided, using a committed leader of a later wave.
It scans the decisions at least one wave above the slot, oldest first, and passes over skipped slots (\cref{line:anchor}). If the first remaining decision is undecided, so is the slot. Otherwise that decision commits a block $A$, the \emph{anchor}. $A$ commits the leader block with at least \plink certificates in its causal history (\cref{line:link}), or skips the slot if no block qualifies. When \plink is low enough that two conflicting blocks of an equivocating leader both qualify, the least identifier breaks the tie. Anchors sit at least a full wave above the slot so that every block there reaches at least $q$ of the slot's decision-round blocks (\cref{sec:fabric:new}).

\subsection{Parameters}
\label{sec:fabric:params}

The \decider has six parameters, which determine how each slot is decided. The parameters are:
\begin{itemize}
      \item \pwavelength: rounds from proposal to decision. The shortest wave has two rounds, proposal and vote. A separate certificate round adds one, so $\pwavelength \ge 3$ when votes are not certificates, and any further rounds are buffer rounds.
      \item \pvoteiscert: whether a vote is its own certificate. When set, the vote round is the decision round and \pcert is unused.
      \item \pcommit: distinct certifying authors in the decision round needed for a direct commit.
      \item \pskip: distinct blaming authors in the vote round needed for a direct skip. Setting $\pskip = n$ disables direct skips because on a DAG with self-parents the leader's own vote-round block never blames its slot.
      \item \pcert: distinct voting authors a decision-round block must reference to certify a leader (used when \pvoteiscert is false).
      \item \plink: distinct certifying authors in the anchor's history needed for an indirect commit.
\end{itemize}

A protocol also fixes the schedule (\pleaders and \ppipeline), the leader election (\pasync), and the DAG's parent quorum $q$.

\para{Schedule and election}
\pleaders is the number of leaders per round, and \ppipeline controls whether a wave starts in every round, so that neighboring waves overlap, or only every \pwavelength rounds. Both parameters specify which slots exist without changing the decision rule, so the proofs treat the schedule separately. Under partial synchrony, \pasync is false and validators follow a public deterministic schedule (such as round-robin, leader reputation~\cite{hammerhead}, or an adaptive schedule, \cref{sec:features}) with the leader timeout of \cref{sec:background}. Under asynchrony, \pasync is true and validators reveal a common coin~\cite{cachin2005random} in the decision round, using buffer rounds to spread proposals before the coin names the leader.

\para{Parent quorum}
The parent quorum $q$ is a DAG parameter that lower-bounds how many decision-round blocks any later block reaches. In the published three-round protocols $q$ equals \pcert, but the two differ in the hybrid three-round rows of \cref{tab:params}, where $\pcert = \lceil (n+f+1)/2 \rceil$ falls below $q = n - f - c$ once $n > 3f+2c+2$. We therefore list $q$ as its own column, as the implementation does (\cref{sec:implementation}), and set $q = n - f - c$ throughout.

\para{Liveness}
Liveness requires thresholds $\pcommit, \pcert \le n - f - c$ (\cref{sec:background}). The skip threshold \pskip is exempt, since the indirect rule decides the slots a direct skip leaves open.

\subsection{Recovering Existing Protocols}
\label{sec:fabric:existing}

\begin{table*}[t]
      \centering\footnotesize\setlength{\tabcolsep}{4pt}
      \caption{Configurations of the \decider. $^{\ast}$No direct skip.}
      \label{tab:params}
      \begin{tabular*}{\textwidth}{@{\extracolsep{\fill}}llccccccccc@{}}
            \toprule
            Protocol             & Ref.              & $n$           & Network & Wave   & Vote is cert. & Commit       & Skip                             & Cert.                          & Link          & $q$         \\
            \midrule
            Cordial Miners       & \cite{cordial}    & $3f{+}1$      & PS      & 3      & \no           & $n{-}f$      & $n^{\ast}$                       & $n{-}f$                        & $1$           & $n{-}f$     \\
            Cordial Miners Async & \cite{cordial}    & $3f{+}1$      & Async   & 5      & \no           & $n{-}f$      & $n^{\ast}$                       & $n{-}f$                        & $1$           & $n{-}f$     \\
            Mysticeti            & \cite{mysticeti}  & $3f{+}1$      & PS      & 3      & \no           & $n{-}f$      & $n{-}f$                          & $n{-}f$                        & $1$           & $n{-}f$     \\
            Mahi-Mahi            & \cite{mahimahi}   & $3f{+}1$      & Async   & 4 or 5 & \no           & $n{-}f$      & $n{-}f$                          & $n{-}f$                        & $1$           & $n{-}f$     \\
            BlueBottle           & \cite{bluebottle} & $5f{+}1$      & PS      & 2      & \yes          & $n{-}f$      & $n{-}f$                          & --                             & $n{-}3f$      & $n{-}f$     \\
            BlueBottle Async     & \cite{bluebottle} & $5f{+}1$      & Async   & 3      & \yes          & $n{-}f$      & $n{-}f$                          & --                             & $n{-}3f$      & $n{-}f$     \\
            Orcaella             & \cite{orcaella}   & $5f{+}3c{+}1$ & PS      & 2      & \yes          & $n{-}f{-}c$  & $4f{+}2c{+}1$                    & --                             & $n{-}3f{-}2c$ & $n{-}f{-}c$ \\
            Nemo-Nemo            & \cite{nemonemo}   & $2c{+}1$      & PS      & 2      & \yes          & $n{-}c$      & $n^{\ast}$                       & --                             & $1$           & $n{-}c$     \\
            \midrule
            Orcaella Async       & \emph{new!}       & $5f{+}3c{+}1$ & Async   & 3      & \yes          & $n{-}f{-}c$  & $4f{+}2c{+}1$                    & --                             & $n{-}3f{-}2c$ & $n{-}f{-}c$ \\
            Nemo-Nemo Async      & \emph{new!}       & $2c{+}1$      & Async   & 3      & \yes          & $n{-}c$      & $n^{\ast}$                       & --                             & $1$           & $n{-}c$     \\
            Hybrid Mysticeti     & \emph{new!}       & $3f{+}2c{+}1$ & PS      & 3      & \no           & $2f{+}c{+}1$ & $\lfloor\tfrac{n+f+1}{2}\rfloor$ & $\lceil\tfrac{n+f+1}{2}\rceil$ & $1$           & $n{-}f{-}c$ \\
            Hybrid Mahi-Mahi     & \emph{new!}       & $3f{+}2c{+}1$ & Async   & 4 or 5 & \no           & $2f{+}c{+}1$ & $\lfloor\tfrac{n+f+1}{2}\rfloor$ & $\lceil\tfrac{n+f+1}{2}\rceil$ & $1$           & $n{-}f{-}c$ \\
            \bottomrule
      \end{tabular*}
\end{table*}

The upper half of \cref{tab:params} lists eight published protocols as configurations of the \decider. We take each row's thresholds and wave length from the protocol's paper, and set \pvoteiscert when the protocol has no separate certificate round. Each row makes the same decisions as its protocol, except possibly where \textsc{Vote} breaks ties among an equivocating leader's blocks. Without equivocation, the decisions are identical. We prove in Lean that each row equals our formalization of its protocol (\cref{sec:proofs}).

\para{Three-round partially synchronous rows}
Mysticeti is the \decider at $\pwavelength = 3$ with every threshold at $n - f$ and a link of one certificate. Two certificate quorums intersect in a correct validator, so a slot has at most one certified block and a single certificate in the anchor's history identifies it. Cordial Miners is the same row without the direct skip, and with one leader per wave and no pipelining. Its published rule skips only through an anchor, which $\pskip = n$ encodes directly on a DAG with self-parents.

\para{Two-round rows}
BlueBottle has no certificate round ($\pvoteiscert$), so its vote round is its decision round. At $n = 5f+1$, a direct commit needs $n - f$ votes and an indirect commit needs $n - 3f$ votes in the anchor's history. Orcaella extends BlueBottle to $f$ Byzantine and $c$ crashed validators at $n = 5f+3c+1$. Its commit and parent quorums are $n - f - c$, its link is $n - 3f - 2c$, and its skip threshold is $4f + 2c + 1$, which equals $n - f - c$ at this bound. Nemo-Nemo commits in two rounds under crash faults alone, at $n = 2c+1$, with commit and parent quorums of $c + 1$, a link of one, and no direct skip. Its published blocks have no self-parent. \sysname adds one, so that $\pskip = n$ disables the direct skip, as in Cordial Miners.

\para{Asynchronous rows}
Mahi-Mahi, BlueBottle Async, and Cordial Miners Async use the thresholds of Mysticeti, BlueBottle, and Cordial Miners, respectively. They set \pasync, so a common coin opened in the decision round elects the leader, and add buffer rounds before the vote. Mahi-Mahi uses one or two buffer rounds ($\pwavelength \in \{4, 5\}$), BlueBottle Async one, and Cordial Miners Async two. With buffer rounds, a voter's history can contain several blocks of an equivocating leader, and different \textsc{Vote} rules can pick different ones.

\subsection{New Configurations}
\label{sec:fabric:new}

The lower half of \cref{tab:params} lists four new protocols. The safety inequalities below fix their thresholds, and the \decider's proofs apply unchanged. All four are safe, and the asynchronous ones decide every slot with probability 1 against an adversary that cannot predict the coin.

\para{Safety inequalities}
Fix $n$ validators with at most $f$ Byzantine, a parent quorum $q$, and a row of the table. Two facts underlie the inequalities. First, two sets of same-round blocks with more than $n + f$ authors in total share a block by a non-Byzantine author, which votes for at most one block of a slot. Second, a block in a later round reaches at least $a + q - n - f$ of the decision-round blocks of any $a$ authors. Each wave shape yields three inequalities:

\begin{align*}
       & \textit{Two rounds (\pvoteiscert)}            &  & \textit{Three rounds}                   \\
       & \mathit{skip} + \mathit{link} \ge n+f+1       &  & 2\,\mathit{cert} \ge n+f+1              \\
       & \mathit{commit} + \mathit{link} \ge n+f+1     &  & \mathit{skip} + \mathit{cert} \ge n+f+1 \\
       & \mathit{link} + n + f \le \mathit{commit} + q &  & \mathit{commit} + q \ge n+f+1           \\
\end{align*}
In both columns, one inequality keeps a direct skip consistent with a later indirect commit ($\mathit{skip} + \mathit{link}$ and $\mathit{skip} + \mathit{cert}$). Another prevents conflicting decisions on an equivocating leader's blocks. In two rounds it keeps a direct commit consistent with an indirect decision on a conflicting block, and in three rounds it makes certificates unique. The last ensures that every anchor reaches enough certificates of a directly committed leader. The three-round family also fixes $\mathit{link} = 1$. If a row meets its column's inequalities, any two views decide every slot the same way, under every vote rule (\cref{app:security}). Every row of \cref{tab:params} meets them except the two Nemo-Nemo rows, which meet their crash-only weakening (below).
\para{Solving for the thresholds}
Add the liveness bound $\mathit{commit}, \mathit{cert} \le n - f - c$ and set $q = n - f - c$. The two-round inequalities have a solution if and only if $n \ge 5f + 3c + 1$. At this bound they force Orcaella's thresholds, and at $c = 0$ BlueBottle's. The three-round inequalities have a solution if and only if $n \ge 3f + 2c + 1$. At this bound they force $\mathit{commit} = 2f + c + 1$ and $\mathit{cert} = \lceil (n+f+1)/2 \rceil$. The smallest skip threshold that $\mathit{skip} + \mathit{cert} \ge n+f+1$ allows is $\lfloor (n+f+1)/2 \rfloor$, and the correct validators alone reach it. We call this configuration \emph{Hybrid Mysticeti}. At $c = 0$ it is Mysticeti. The bounds differ because two-round rules count votes, which conflicting blocks of an equivocating leader can both collect, so $\mathit{link}$ must exclude a conflicting block yet stay reachable from every anchor. A certificate round makes certificates unique, so one certificate suffices, at the cost of a round.
For $f = 0$, no validator equivocates, so a slot has at most one block and the inequality against conflicting blocks drops out. The remaining two-round inequalities admit $n \ge 3c+1$ with a direct skip ($\pskip \le n - c$) and $n \ge 2c + 1$ without one ($\pskip = n$). Nemo-Nemo is the second case, and omitting the direct skip reduces its committee from $3c+1$ to $2c+1$.

\para{Asynchronous configurations}
Setting \pasync and adding buffer rounds to Orcaella, Nemo-Nemo, and Hybrid Mysticeti gives \emph{Orcaella Async}, \emph{Nemo-Nemo Async}, and \emph{Hybrid Mahi-Mahi}. The inequalities do not depend on \pwavelength, so safety is unchanged. Liveness depends on how many leaders are \emph{safe to commit} (\emph{safe} for short) before the coin opens: every valid vote-round block reaches the leader's proposal, so if the coin names that leader, its slot commits whatever the scheduler does next. The scheduler acts before the coin is revealed, so it cannot single out the leader the coin will name, only limit how many leaders are safe. Every asynchronous row has $2q \ge n + f + 1$, which gives a common core after two rounds. One buffer round then guarantees at least one safe leader, and two guarantee at least $q - f$. Further buffer rounds do not help: an adversarial scheduler can hold the count at $q - f$.
When $2q = n + f + 1$, as at $n = 3f+1$, $3f+2c+1$, and $2c+1$, one buffer round guarantees at least one safe leader, so the coin names a safe leader with probability at least $1/n$. Mahi-Mahi, Cordial Miners Async, and Hybrid Mahi-Mahi therefore use up to two buffer rounds. Nemo-Nemo Async uses one buffer round and guarantees at least one safe leader, as do Mahi-Mahi and Hybrid Mahi-Mahi at $\pwavelength = 4$. At $n \ge 5f+3c+1$, one buffer round guarantees at least $n/3$ safe leaders, so the coin names a safe leader with probability at least $1/3$. BlueBottle Async and Orcaella Async therefore use one buffer round.
With more than $n - s$ leaders per round, where $s$ is the number of safe leaders, at least one leader of the round is safe.

All four configurations are implemented in our library (\cref{sec:implementation}) and evaluated alongside the published protocols (\cref{sec:evaluation}). Hybrid Mysticeti and Hybrid Mahi-Mahi tolerate $f$ Byzantine and $c$ crashed validators with $3f + 2c + 1$ validators, compared to Orcaella's $5f + 3c + 1$, at the cost of one more round. Orcaella Async and Nemo-Nemo Async bring the two-round hybrid and crash rules to asynchrony with one buffer round. The two differ in how quickly they decide: Orcaella Async keeps at least $n/3$ safe leaders, so the coin names one with probability at least $1/3$ per wave, whereas Nemo-Nemo Async guarantees at least one, a probability of at least $1/n$. Both decide every slot with probability 1 (\cref{lem:validity-async}), but against an adversarial scheduler Nemo-Nemo Async may need on the order of $n$ waves, trading worst-case latency for one fewer buffer round.

\subsection{Fault Tolerance Versus Latency}
\label{sec:fabric:tradeoff}

\Cref{tab:params} shows the trade-off between fault tolerance and latency. Every row builds the same DAG, one block per validator per round with $q$ parents each, so throughput is the same and rows differ only in latency, measured in rounds. Three choices set this latency. First, for a given wave length, a stronger adversary needs a larger committee. Two-round waves need $5f + 3c + 1$ validators and three-round waves $3f + 2c + 1$, so at a fixed committee the three-round rule tolerates more faults and the two-round rule commits one round sooner. Second, asynchrony adds buffer rounds so proposals spread before the coin names the leader, one at $5f+1$, $5f+3c+1$, and $2c+1$, and up to two at $3f+1$ and $3f+2c+1$. Third, omitting the direct skip, as Cordial Miners and Nemo-Nemo do, adds no latency without faults, but leaves every crashed leader to the indirect rule, one wave later.
Without faults, commit latency is \pwavelength rounds, plus the wait for a leader slot without pipelining (\cref{sec:evaluation}).
\section{Layering Mechanisms Around the \decider}
\label{sec:features}

The \decider algorithm is the heart of our consensus protocol as it achieves the hardest and most critical part, the total ordering. \Cref{alg:decider} is only 16 lines of pseudocode. Its simplicity, however, is not sufficient for the needs of a dynamic, heterogeneous environment such as a production blockchain.
We take Sui as an example. Sui deploys additional mechanisms along Mysticeti: (a) responsive liveness, so that rounds advance at network speed rather than on timeouts, (b) a leader scheduling algorithm that adapts slot assignments to validators' varying performance, (c) a crash recovery mechanism for each validator, which is outside the classic BFT adversary, and (d) a checkpoint tier that exposes higher resilience guarantees to light clients at the cost of one extra message delay.  Each of these mechanisms is as critical and as error-prone as the core consensus algorithm.  In 3\,819 commits to Sui's consensus, checkpoint, and execution code we count 96 bug fixes related to them. Taking a holistic approach, \sysname provably composes each mechanism with the \decider as well as with the other mechanisms.

\para{How the mechanisms compose}
Every mechanism reads the \decider only through two properties: decisions agree across views, and a correctly led slot commits once its wave is built and certified (\cref{sec:proofs}). Composing a mechanism with any configuration, or with the other mechanisms, therefore needs no new argument about the \decider. All four mechanisms compose under partial synchrony, and all but responsive liveness under asynchrony.

\para{Mechanism 1: Responsive liveness}
A responsive validator builds its next block as soon as it holds what the \decider requires: a block reaching the leader's block or the votes certifying it. It waits out the timeout only as a fallback. Rounds thus usually advance at the network's actual latency rather than at worst-case delay bounds.

\para{Mechanism 2: Leader scheduling}
Real blockchains reassign leader slots to validators they have observed to perform well: HammerHead~\cite{hammerhead} assigns more leader slots to the validators whose blocks gathered the most votes under the schedule in force. Barnacle~\cite{barnacle} raises the number of leaders per round by one while the fraction of directly decided slots stays high and halves it when the fraction drops. Both read the agreed DAG to change the schedule, yet the schedule determines which slots exist and so what the \decider decides. Unless the two are stratified, agreement is circular. \sysname runs both as one mechanism built on \emph{schedule configurations}. A schedule configuration fixes the slots of each round, their leaders, and an interval, the range of rounds it governs, and its slots are decided under the configuration's schedule, reading slots past the interval only as anchors. The first committed slot past the interval is the \emph{pivot}, and an update rule reads the pivot's causal history and emits the next schedule configuration, which takes effect after the pivot's round (Barnacle) or at the end of the interval, re-deciding the slots in between (HammerHead). Nothing in a schedule configuration's range depends on the next one, so the dependency is well-founded. Safety holds for any update rule that reads only the pivot's history, with no synchrony or fairness assumption, whereas liveness needs schedules that keep committing (\cref{sec:proofs}).

\para{Mechanism 3: Checkpoints}
Like Orcaella~\cite{orcaella}, \sysname adds a second tier of finality, but to any \decider. Validators sign the state they reach by executing what their \decider commits. A quorum of signatures on one state is a \emph{checkpoint certificate}, and a quorum of acknowledgements of a certificate makes the checkpoint \emph{final}. Two certificates of one epoch and height bind one checkpoint, even when the inputs contain forks. The checkpoint tier finalizes every slot the \decider commits, under leader scheduling, and for responsive validators that crash and restart.

\para{Mechanism 4: Crash recovery}
A validator writes every block it receives or creates to disk before acting on it. On restart, it rebuilds every round it missed and asks its peers again for what it lacks. It never builds two blocks for the same round, so safety is unchanged. Once the network has stabilized and the correct validators have stopped crashing, the run behaves like a crash-free run whose network stabilizes at that point, so the liveness results of \cref{sec:proofs} apply from then on. Under asynchrony, Agreement holds and Validity holds with probability $1$. This covers all validators crashing and restarting, which matters in real-world blockchains: honest validators routinely update their software and reboot while the network is running, with no observable downtime. Occasionally they all need to apply an emergency fix simultaneously to mitigate a live vulnerability. Crash recovery lets the network resume after such a simultaneous restart.

\section{Formal Proofs}
\label{sec:proofs}

We formalize \sysname in Lean~4~\cite{leandag,lean4,mathlib}
and prove that every configuration of the \decider implements atomic broadcast against its own fault and network model, and that the mechanisms of \cref{sec:features} preserve it. We also formalize the eight published protocols and show that their commit rules are equivalent to configurations of the \decider. The definitions and proofs are about 125{,}000 lines of Lean, 40{,}000 of them concrete executions that exercise every definition. The \decider's theorems are stated in files that contain no proofs, and every principal result rests on Lean's three standard axioms. This split reflects how the development was built: AI agents produced most of the proofs, which Lean's kernel checks, while the definitions and theorem statements, the part a reader must trust, are kept separate and reviewed by hand. \Cref{app:security} gives the proofs of \cref{thm:abc,thm:compose} in more detail.

\para{Atomic broadcast}
We prove the definition of \cref{sec:background} in the two forms of the literature, Sailfish's Definition~1~\cite{sailfish} and Sui et al.'s Definition~2.1~\cite{sui-duan}. Agreement and Validity hold for the \emph{reliable} validators: correct validators that eventually receive every block another reliable validator holds. The adversary controls message delays but does not see the common coin. Stake and the deduplication of transactions are not modeled.

\begin{theorem}[Atomic broadcast]
    \label{thm:abc}
    Every configuration of the \decider in \cref{tab:params} implements atomic broadcast (\cref{sec:background}) against its own fault and network model, for the reliable validators and under the leader schedules of \cref{lem:validity-ps}.
\end{theorem}
Only Validity depends on the network model. Prefix consistency, total order, and integrity hold in every execution, whereas agreement additionally requires eventual delivery among the reliable validators.
Under partial synchrony, after GST, Validity holds under round-robin in every configuration of the \decider, and under any leader schedule with what we call \emph{bounded anchor chains} (\cref{lem:validity-ps}). Under asynchrony, the common coin retroactively schedules leaders, and as long as the adversary cannot predict it, every slot commits with a probability that approaches 1 as the run grows (\cref{lem:validity-async}).
\begin{theorem}[Composition]
    \label{thm:compose}
    At every configuration of the \decider the mechanisms of \cref{sec:features} compose (all four under partial synchrony and all but responsive liveness under asyncrhony) and implement atomic broadcast against the configuration's fault model.
\end{theorem}

\section{Implementation}
\label{sec:implementation}

We implement \sysname in Rust~\cite{dugong-code} on top of the open-source Mysticeti codebase~\cite{mysticeti}. %
The implementation comprises roughly 25k lines of Rust across seven crates, including the DAG engine (blocks, round advancement, storage, networking, and synchronization), the \decider, the replica, a discrete-event simulator, and a command-line interface.

\para{Unified \decider}
The \decider implements a single commit rule shared across all protocols. Each protocol is specified as a flat configuration structure holding the parameters of \cref{sec:fabric} and two implementation fields: the round-advancement threshold and whether validators wait for leader blocks. In the Rust code, the \decider instantiates one committer per pipelined wave offset and leader slot, evaluates the direct rule followed by the indirect rule on each slot, and outputs the longest decided prefix of leaders. The commit logic contains no protocol-specific branches: each of the twelve configurations evaluated in \cref{sec:evaluation} is simply a constructor that checks the committee size bound and sets the corresponding parameters. When $f = 0$, signature generation and verification are disabled, since crashed validators do not forge messages.

\para{Simulator}
The simulator runs the unmodified consensus code, including the DAG engine, the \decider, the replica logic, and storage, and replaces only the asynchronous runtime: a thin abstraction swaps Tokio for a deterministic discrete-event scheduler that controls virtual time, task spawning, and message delivery, so that a single random seed determines the execution. Because all simulated validators run on one machine, CPU contention is not representative of a distributed deployment, so we stub out cryptographic operations. Point-to-point links are FIFO with a bounded window of in-flight messages, and per-message latency is drawn either uniformly or from measured inter-region round-trip times, with optional exponential jitter. Faults are injected at the network layer (\cref{sec:eval:setup}).
Asynchronous configurations elect leaders via a seeded pseudo-random function instead of a common coin.

\section{Evaluation}
\label{sec:evaluation}

We evaluate all twelve configurations of the \decider using the same codebase and simulated network environment. We test four claims:
\begin{itemize}
      \claimitem{claim:c1} Fault-free commit latency is determined by the number of rounds a configuration requires to commit, rather than by its fault model or quorum size.
      \claimitem{claim:c2} Every configuration remains live up to its theoretical fault bound. Under crash faults, the random leader election of asynchronous configurations avoids bursts of consecutive crashed leaders, reducing tail latency at the expense of a higher median latency.
      \claimitem{claim:c3} An equivocating leader delays its slot by roughly one wave and blocks subsequent slots, which penalizes long, pipelined waves the most.
      \claimitem{claim:c4} Under high network jitter, partially synchronous protocols with a separate certificate round degrade sharply, but two-round protocols and asynchronous protocols with a single buffer round maintain high direct commit rates.
\end{itemize}

Our experiments evaluate protocol performance under crash faults and leader equivocation. Benchmarking BFT protocols against arbitrary Byzantine strategies is an open research question~\cite{twins}, and we establish worst-case safety and liveness through our mechanized proofs (\cref{sec:proofs}).

\subsection{Experimental Setup}
\label{sec:eval:setup}

We denote hybrid configurations by their fault bounds $(f, c)$ and test three operating points for each family: \emph{balanced}, \emph{Byzantine-heavy}, and \emph{crash-heavy}. Specifically, Hybrid Mysticeti and Hybrid Mahi-Mahi are configured at $(9, 9)$, $(13, 3)$, and $(2, 19)$, while Orcaella and Orcaella Async are configured at $(6, 5)$, $(7, 3)$, and $(2, 11)$, matching the ratios of the Orcaella paper~\cite{orcaella}. The last two exceed Orcaella's minimum committee and so run a safe variant with a lower skip threshold (\cref{sec:fabric:new}).

We simulate a committee of $n = 46$ validators, a network size comparable to prior DAG consensus evaluations~\cite{narwhal,bullshark,mysticeti,mahimahi,sailfish}. Since $46 = 3 \cdot 15 + 1 = 5 \cdot 9 + 1$, both the $3f+1$ and $5f+1$ resilience bounds are tight, and every hybrid bound is met within at most two validators. Link latencies are drawn uniformly between 40 and 80\,ms. We denote the average duration of a DAG round by $\rho$, which is approximately 65\,ms without faults or jitter. Because absolute latencies scale with the underlying link delay, we also report latency differences in multiples of $\rho$. Partially synchronous configurations use a leader timeout of 160\,ms, and each validator generates transactions at a constant rate across all runs. When network jitter is enabled, each message experiences an additional delay drawn from an exponential distribution with a mean of 120\,ms, capped at 640\,ms. Each simulation runs for 120\,s, and metrics are collected after a 30\,s warm-up period across all live validators. We measure end-to-end latency from the time a transaction is submitted until it is committed.

To stress-test the protocols, we assign faults to consecutive validator indices: crashed validators occupy the last $k$ indices and equivocating validators occupy the first $e$ indices, so a round-robin schedule encounters all faulty leaders consecutively. Crashed validators are disconnected at the start of the run. Following production implementations, validators skip the leader timeout if their connection to the leader is down, so crashed leaders do not trigger timeout delays (safe, but outside the liveness proofs).
In each round where it holds a leader slot, an equivocating validator sends every peer two conflicting blocks with the same content and different digests, in opposite order to odd- and even-indexed peers, so half of the committee sees each block first~\cite{twins}.

For each data point, we report the median across 7 random seeds of the p50 latency, with error bars extending to the p90 latency. Across all $\sim$3,000 runs of our evaluation campaign (roughly 2,500 vCPU-hours), the simulator's end-of-run safety check never found two honest validators that committed conflicting sequences.

\subsection{Fault-Free Latency}
\label{sec:eval:happy}

\begin{figure}[t]
      \centering
      \includegraphics[width=\columnwidth]{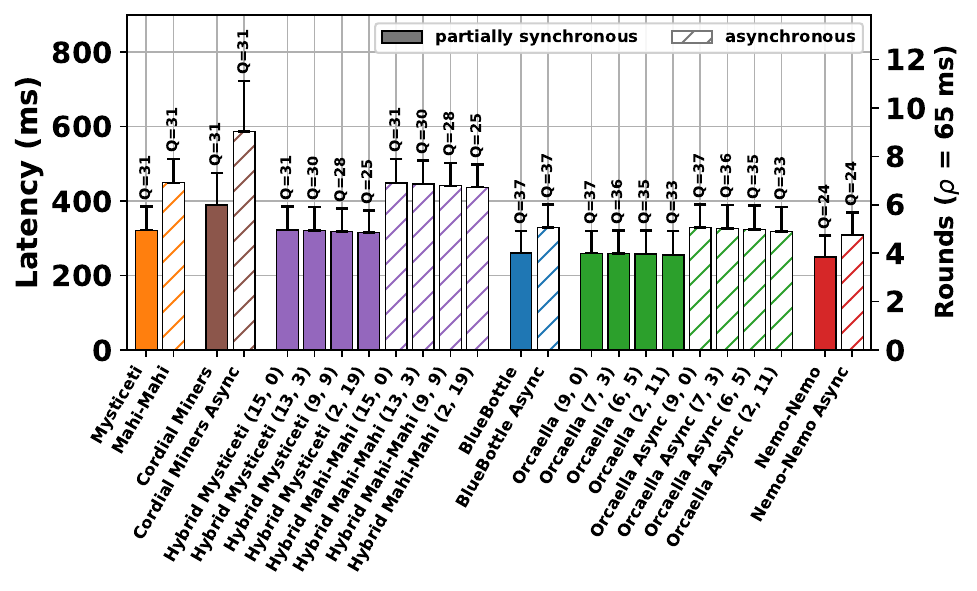}
      \caption{Fault-free latency of every protocol family (p50, whisker to p90; solid: partially synchronous, hatched: asynchronous). Hybrid configurations are shown at all four $(f, c)$ points, from $c = 0$ to crash-heavy, and $Q$ is the round quorum. Right axis: rounds of $\rho = 65$\,ms.}
      \label{fig:eval:happy}
\end{figure}

Protocols cluster by the number of DAG rounds required to commit rather than by their fault model (\cref{fig:eval:happy}). In particular, all two-round partially synchronous protocols achieve nearly identical p50 latency (around 255\,ms), whether they tolerate crash faults (Nemo-Nemo), Byzantine faults (BlueBottle), or hybrid faults (Orcaella). Similarly, the three-round partially synchronous protocols, Mysticeti and Hybrid Mysticeti, commit approximately $1\rho$ later.

The round quorum size $Q$ barely matters. Within the hybrid families, moving from $(15, 0)$ to $(2, 19)$ reduces Hybrid Mysticeti's quorum from 31 to 25 while changing its p50 latency by only $0.1\rho$, and moving from $(9, 0)$ to $(2, 11)$ reduces Orcaella's quorum from 37 to 33 with the same $0.1\rho$ difference. The asynchronous hybrid families vary by at most $0.2\rho$. The same holds across protocols: Nemo-Nemo ($Q = 24$) commits within $0.2\rho$ of BlueBottle ($Q = 37$). At $c = 0$, each hybrid protocol matches the latency of its corresponding pure-Byzantine baseline (Orcaella matches BlueBottle, Hybrid Mysticeti matches Mysticeti, and Hybrid Mahi-Mahi matches Mahi-Mahi) within 1\,ms.

Each asynchronous protocol adds latency proportional to its buffer rounds: the three-round asynchronous protocols (BlueBottle Async, Orcaella Async, and Nemo-Nemo Async) commit $1\rho$ later than their partially synchronous counterparts, and the five-round protocols (Mahi-Mahi and Hybrid Mahi-Mahi) commit $2\rho$ later, while the common coin itself introduces no measurable overhead. The only exceptions are the unpipelined Cordial Miners variants: because Cordial Miners elects a leader only once per $w$-round wave, transactions wait an average of $(w - 1)/2$ additional rounds before entering a leader wave, placing Cordial Miners $1\rho$ above Mysticeti and Cordial Miners Async $2\rho$ above Mahi-Mahi. These observations validate \cref{claim:c1}.
\takeaway{1}{Fault-free latency is governed by the number of rounds in a wave: three-round rules commit $1\rho$ after two-round rules, and each asynchronous buffer round adds $1\rho$, whereas changing the quorum size shifts latency by at most $0.2\rho$.}

\subsection{Crash Faults}
\label{sec:eval:crashes}

\begin{figure*}[t]
      \centering
      \begin{minipage}{0.49\textwidth}
            \centering
            \includegraphics[width=\textwidth]{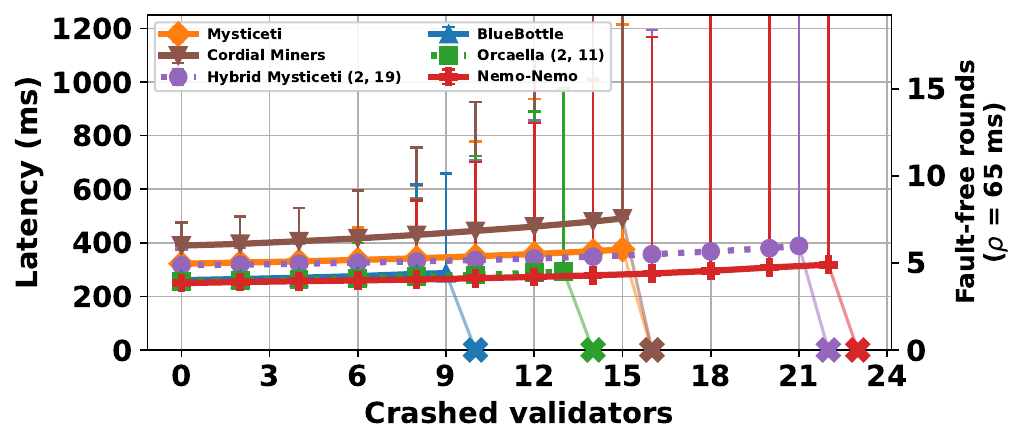}
      \end{minipage}\hfill
      \begin{minipage}{0.49\textwidth}
            \centering
            \includegraphics[width=\textwidth]{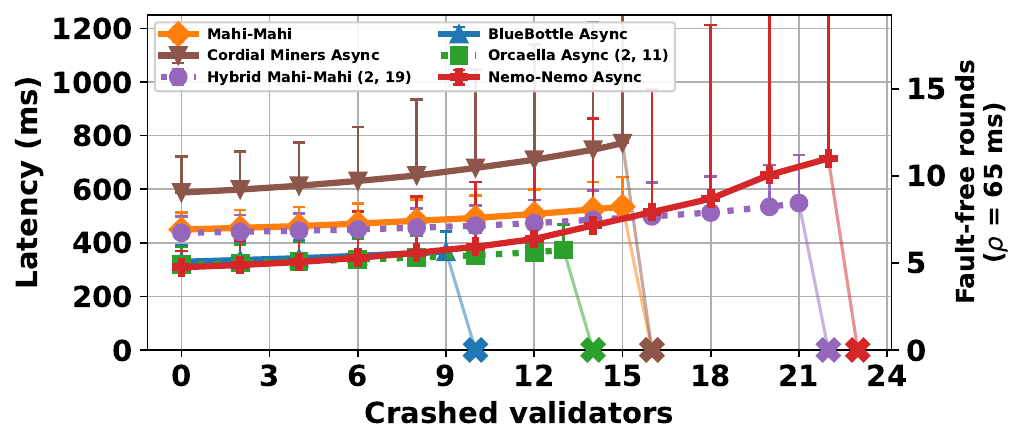}
      \end{minipage}
      \caption{Latency as validators crash, for partially synchronous (left) and asynchronous (right) configurations (p50, whisker to p90). An X marks the first crash count at which a configuration stalls. Hybrid configurations are shown at their crash-heavy point. Right axis: fault-free rounds of $\rho = 65$\,ms. Rounds lengthen to about 79\,ms at each fault bound. Whiskers above 1,250\,ms are clipped.}
      \label{fig:eval:crashes}
\end{figure*}

\cref{fig:eval:crashes} shows latency as the number $k$ of crashed validators grows. Every configuration remains live up to its fault bound and stalls one crash beyond it: 9 crashes for BlueBottle, 13 for Orcaella $(2, 11)$, 15 for Mysticeti, Mahi-Mahi, and Cordial Miners, 21 for Hybrid Mysticeti and Hybrid Mahi-Mahi $(2, 19)$, and 22 for Nemo-Nemo.

\para{Partially synchronous configurations}
Crash faults affect performance in two ways: they reduce the number of active voters and they introduce rounds with crashed leaders. With fewer live validators available to form a quorum of size $Q$, each validator must wait for blocks over slower network links before advancing to the next round. This longer round accounts for the rise in median latency: Mysticeti's p50 stays at $4.7\rho$ while $\rho$ grows from 68.7\,ms to 79.0\,ms at its fault bound. In contrast, crashed leaders drive tail latency. Because crashed validators occupy consecutive indices, the round-robin schedule encounters a burst of $k$ consecutive crashed leaders once every 46 rounds. During this burst, no new leader is committed until the first live leader proposes, which delays transactions submitted during that window (increasing p90 latency) while leaving the rest of the schedule unaffected (preserving p50 latency). Each crashed leader in the burst adds roughly one round to the tail: at the fault bound, the p90 latency of every partially synchronous protocol grows by about $1\rho$ per crashed validator (reaching $13\rho$ for Mysticeti at $k = 15$).

\para{Asynchronous configurations}
Although crashed validators still occupy consecutive indices, the common coin selects each slot's leader uniformly at random. As a result, crashed leaders are distributed across the execution rather than concentrated in a single burst, allowing the protocol to recover within a single wave. For Mahi-Mahi, p90 latency increases by only $2\rho$ at $k = 15$. By $k = 8$, the asynchronous protocols already exhibit lower tail latency than their partially synchronous counterparts, and at the fault bound their p90 latency is roughly cut in half (for example, 727\,ms for Hybrid Mahi-Mahi compared to 1,687\,ms for Hybrid Mysticeti at $k = 21$).

The median latency of the asynchronous protocols, however, stays above that of their partially synchronous counterparts by the number of buffer rounds at every $k$, since the buffer rounds slow down like any other ($2\rho$ between Mahi-Mahi and Mysticeti at $k = 15$, and likewise for the hybrid pair at $k = 21$).

Nemo-Nemo Async and Cordial Miners Async are the only asynchronous configurations whose median latency increases substantially with $k$, growing by about $6\rho$ at $k = 22$ and $3\rho$ at $k = 15$, respectively, compared to $1.3\rho$ for Mahi-Mahi. Both protocols omit the direct skip rule ($S = n$). Consequently, a slot led by a crashed validator cannot be skipped within its own wave and must wait to be resolved indirectly by a later committed anchor, holding back all subsequent slots in the meantime. Random leader election therefore provides no tail benefit for these two protocols, and the p90 latency of Nemo-Nemo Async matches that of partially synchronous Nemo-Nemo. These results confirm \cref{claim:c2}.

\takeaway{2}{All configurations remain live up to their exact fault bounds. Under partial synchrony, crashes increase median latency only by slowing down rounds, while consecutive crashed leaders increase tail latency by about $1\rho$ per crash. Random leader election in asynchronous protocols breaks up bursts of crashed leaders and halves tail latency at the fault bound, provided crashed slots can be skipped directly.}

\subsection{Byzantine and Hybrid Faults}
\label{sec:eval:byzantine}

\begin{figure*}[t]
      \centering
      \begin{minipage}{0.49\textwidth}
            \centering
            \includegraphics[width=\textwidth]{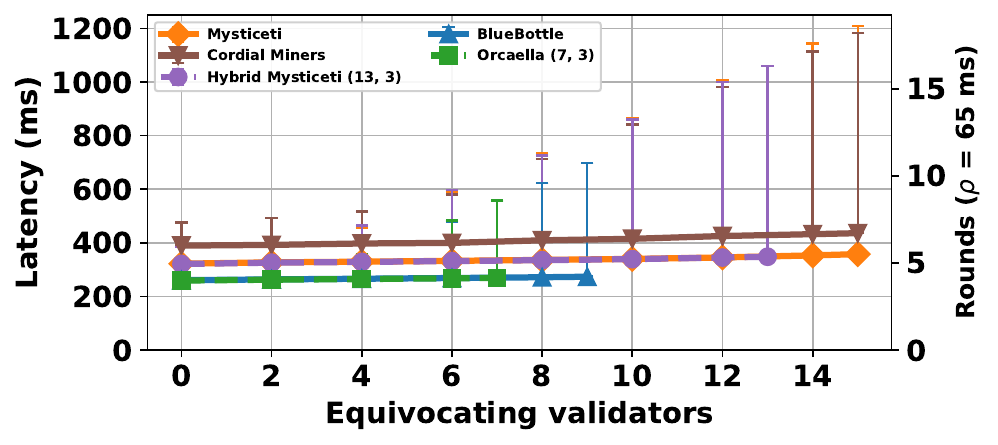}
      \end{minipage}\hfill
      \begin{minipage}{0.49\textwidth}
            \centering
            \includegraphics[width=\textwidth]{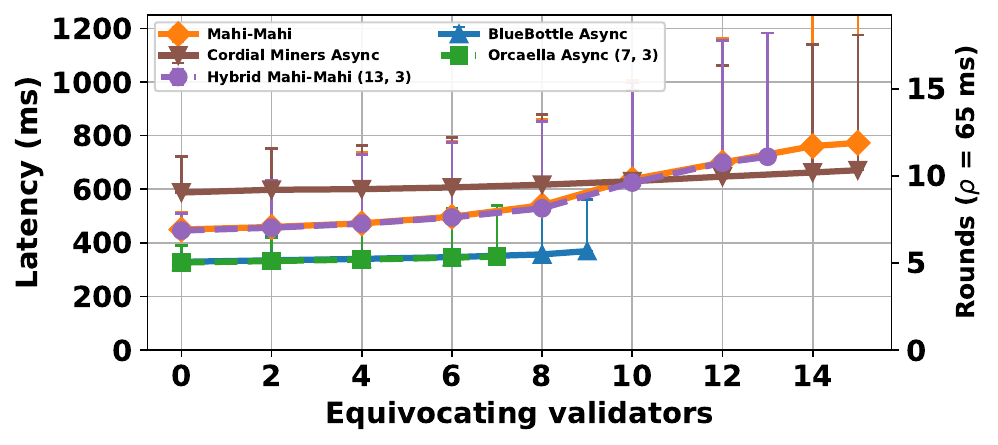}
      \end{minipage}
      \caption{Latency as validators equivocate, with no crashes, for partially synchronous (left) and asynchronous (right) configurations (p50, whisker to p90). Hybrid configurations are shown at their Byzantine-heavy point. Right axis: rounds of $\rho = 65$\,ms. Whiskers above 1,250\,ms are clipped.}
      \label{fig:eval:byzantine}
\end{figure*}

\begin{figure*}[t]
      \centering
      \begin{minipage}{0.49\textwidth}
            \centering
            \includegraphics[width=\textwidth]{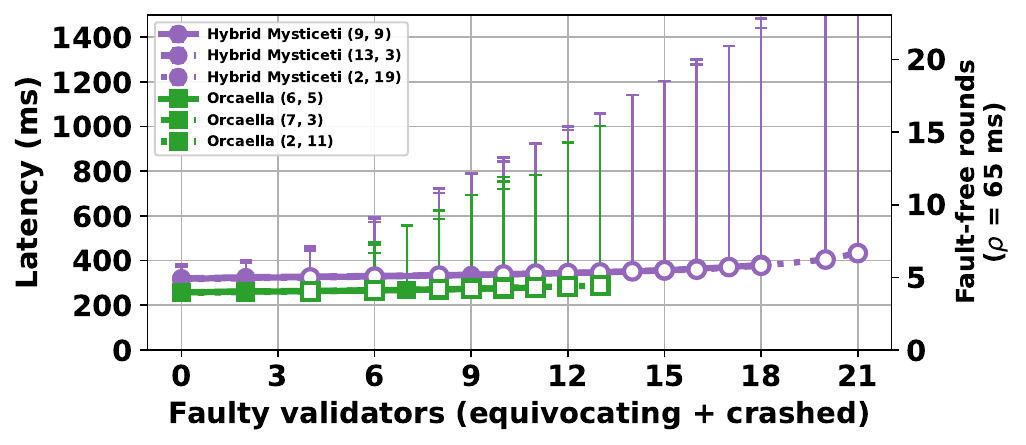}
      \end{minipage}\hfill
      \begin{minipage}{0.49\textwidth}
            \centering
            \includegraphics[width=\textwidth]{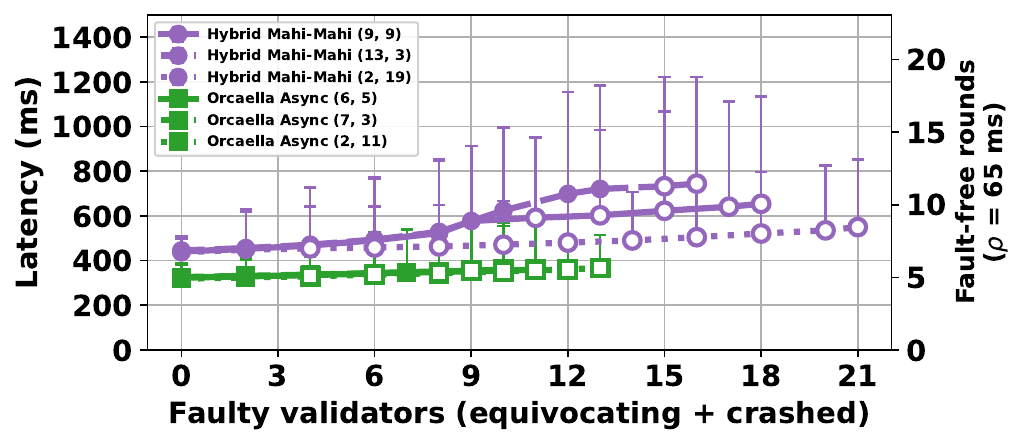}
      \end{minipage}
      \caption{Latency of the hybrid configurations against the total number of faulty validators, for partially synchronous (left) and asynchronous (right) configurations (p50, whisker to p90). Each line first adds equivocators up to $f$ (filled markers), then crashes up to $c$ (hollow markers). Line style marks the $(f, c)$ point: solid balanced, dashed Byzantine-heavy, dotted crash-heavy. Right axis: fault-free rounds of $\rho = 65$\,ms. Whiskers above 1,500\,ms are clipped.}
      \label{fig:eval:hybrid}
\end{figure*}

\cref{fig:eval:byzantine} evaluates performance as $e$ validators equivocate in every round without crashing. When an equivocating validator is elected leader, it splits the committee's votes between its two conflicting blocks, preventing the slot from reaching either the direct commit quorum or the direct skip quorum~\cite{twins}. The fraction of directly committed leaders therefore drops to $1 - e/n$ (0.68 for Mysticeti at $e = 15$), and each equivocating slot can only be resolved indirectly by an anchor at least one wave later, delaying the slot by roughly $w$ rounds and blocking the $w - 1$ slots behind it.

\para{Equivocators only}
For partially synchronous protocols, the latency curves resemble those under crash faults: at $e = 15$, Mysticeti's p50 latency increases by $0.6\rho$ while its p90 latency increases by about $13\rho$. Among the three-round asynchronous protocols, random leader election again yields a slightly higher p50 latency in exchange for a lower p90 latency (for BlueBottle at $e = 9$, the asynchronous p90 is $2\rho$ lower than the partially synchronous one). In contrast, the five-round asynchronous protocols experience a sharp increase in median latency as well: Mahi-Mahi's p50 grows by $5\rho$ at $e = 15$, and Hybrid Mahi-Mahi $(13, 3)$ exhibits the same trend.

Near $e = 10$, Mahi-Mahi's latency exceeds that of Cordial Miners Async because pipelining becomes a disadvantage under frequent equivocation. Mahi-Mahi schedules a leader in every round, so when $e/n = 1/3$ of leaders equivocate and stall for a five-round wave, nearly every transaction is queued behind an undecided slot. Cordial Miners Async, by contrast, elects only one leader per five-round wave, so an equivocating leader delays only its own wave. The two curves cross at the point where this head-of-line blocking outweighs Mahi-Mahi's $2\rho$ fault-free advantage.

\para{Hybrid faults}
\cref{fig:eval:hybrid} first introduces $f$ equivocating validators and then adds up to $c$ crashed validators. Under partial synchrony, the three $(f, c)$ configurations of Hybrid Mysticeti and Orcaella follow nearly the same trajectory, because replacing crashes with equivocators produces two opposing effects: crashed validators stop voting and lengthen every round, whereas equivocating validators continue voting and keep rounds fast, but an equivocating leader stalls its slot until the next anchor instead of being directly skipped. At the same total fault count, an equivocator adds 1--$2\rho$ more to the p90 latency than a crash. Under asynchrony, where the common coin scatters faulty leaders, the three Hybrid Mahi-Mahi configurations separate clearly, with the crash-heavy configuration achieving the lowest latency (at 16 total faults, its p50 is $3.7\rho$ below the Byzantine-heavy configuration) because crashed leaders are directly skipped within their wave while equivocating leaders wait for a later anchor. This confirms \cref{claim:c3}.
\takeaway{3}{An equivocating leader prevents both direct commit and direct skip, stalling its slot for a full wave and blocking subsequent slots. Protocols with waves of up to three rounds keep p50 latency within $1\rho$ at $f$ equivocators, whereas five-round pipelined protocols suffer severe head-of-line blocking and eventually fall behind unpipelined waves.}

\begin{figure*}[t]
      \centering
      \begin{minipage}{0.49\textwidth}
            \centering
            \includegraphics[width=\textwidth]{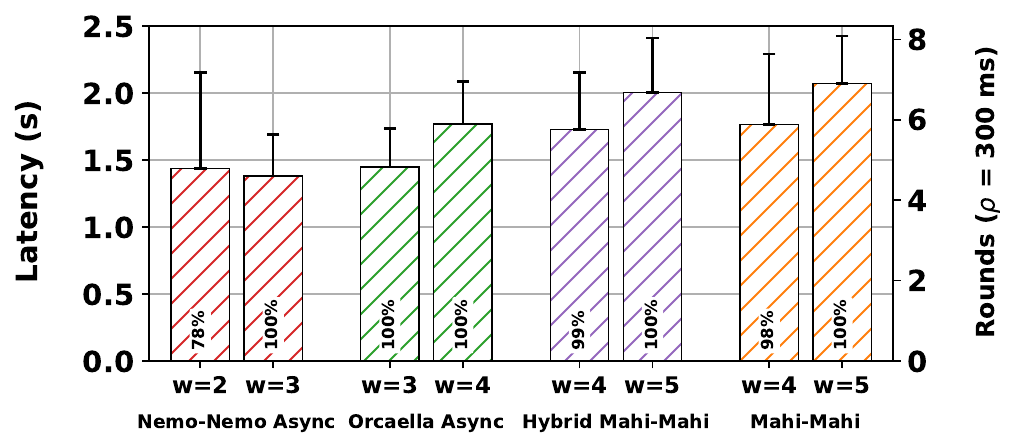}
      \end{minipage}\hfill
      \begin{minipage}{0.49\textwidth}
            \centering
            \includegraphics[width=\textwidth]{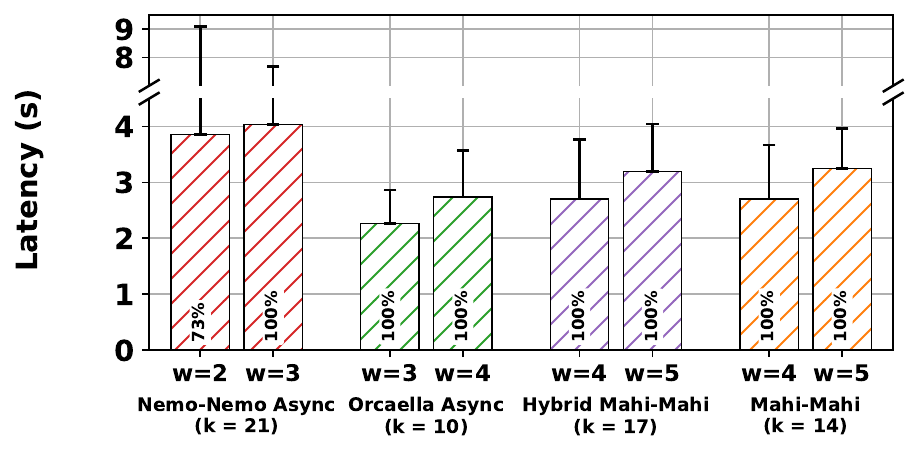}
      \end{minipage}
      \caption{Latency of each asynchronous family at two wave lengths $w$ under jitter, without (left) and with (right) crashes, $k$ one below each fault bound (p50, whisker to p90). The right panel is broken at 4.5\,s. The number at the foot of each bar is its direct commit rate. Right axis (left panel): rounds of $\rho \approx 300$\,ms.}
      \label{fig:eval:buffer}
\end{figure*}

\subsection{Network Jitter}
\label{sec:eval:jitter}

\begin{figure}[t]
      \centering
      \includegraphics[width=\columnwidth]{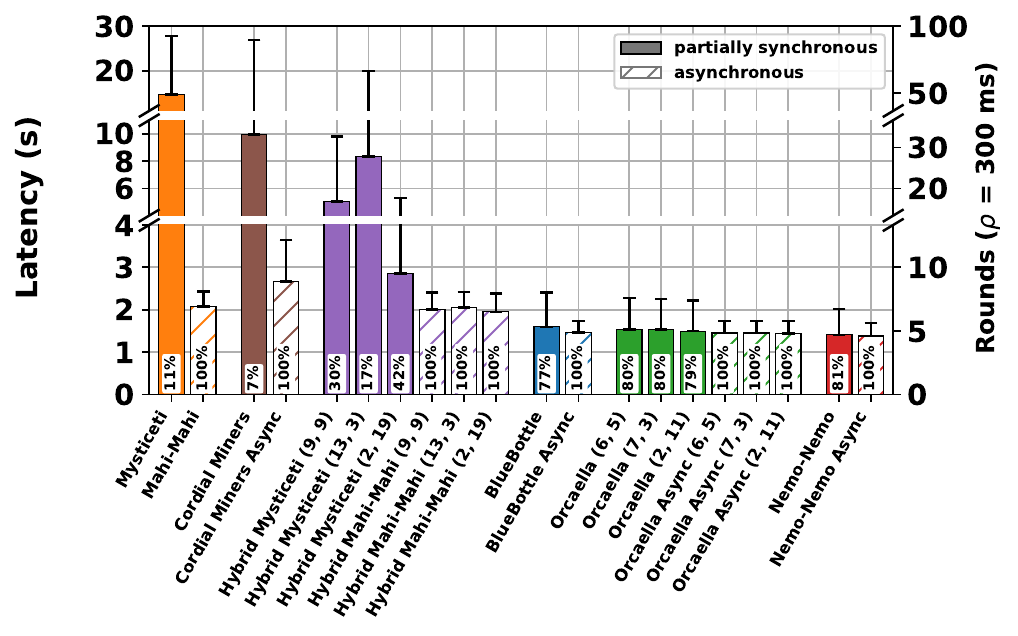}
      \caption{Latency under message jitter without faults (p50, whisker to p90). The axis is broken at 4 and 11\,s. The number at the foot of each bar is its direct commit rate. Right axis: rounds of $\rho \approx 300$\,ms.}
      \label{fig:eval:jitter}
\end{figure}

\cref{fig:eval:jitter} introduces exponential message jitter without validator faults, increasing the average round duration to $\rho \approx 300$\,ms. The protocols fall into three groups. First, all asynchronous configurations continue to commit nearly every leader directly (at least 98\% across all runs, and 100\% in 65 of the 70 runs), so their latency increases only because each round is longer: Mahi-Mahi commits at a p50 of $6.9\rho$, matching its round count without jitter. Second, the two-round partially synchronous protocols (BlueBottle, Orcaella, and Nemo-Nemo) directly commit 77--81\% of their leaders and require roughly one additional round at the median ($4.7$--$5.3\rho$ compared to $3.9\rho$ without jitter). Third, the three-round partially synchronous protocols, which require a separate certificate round (Mysticeti, Cordial Miners, and Hybrid Mysticeti), degrade severely: their direct commit rate drops to 7--42\%, and their p50 latency rises to 10--50 rounds. The three-round asynchronous protocols do not suffer this degradation because they use a two-round commit rule preceded by one buffer round rather than a certificate round. The most affected configurations (Mysticeti, Cordial Miners, and Hybrid Mysticeti $(13, 3)$) do not stabilize within the 120\,s window, with goodput ratios of 0.4 to 1.2 across seeds, so their bars reflect order-of-magnitude behavior.

\para{Impact on the certificate round}
Under a three-round commit rule, a direct commit must assemble a quorum twice: each block in round $r + 2$ must reference $Q$ votes from round $r + 1$ to form a certificate, and the validator must collect $Q$ such certificates. In contrast, a two-round rule requires only $Q$ votes in round $r + 1$. Under jitter, enough blocks in round $r + 1$ still vote for the leader (70--83\% across all configurations), which suffices for a two-round rule. However, because validators advance to round $r + 2$ as soon as they receive $Q$ blocks from round $r + 1$, a typical round-$(r + 2)$ block references only about 36 blocks from round $r + 1$, of which roughly 29 are votes, two short of Mysticeti's threshold $Q = 31$. Consequently, each slot accumulates an average of only 18 certificates instead of the required 31, yielding a direct commit rate of 0.11 (matching the 0.12 rate predicted from the per-round vote distribution). The Hybrid Mysticeti configurations collect a similar number of certificates per slot but require fewer certificates per slot, $2f + c + 1 = 24$ to $30$ instead of 31, which accounts for the $3\times$ spread in their latencies.

Two-round $5f+1$ protocols also differ from Mysticeti in their parent quorum size (37 parents per block for BlueBottle versus 31 for Mysticeti). In \cref{app:ablation}, we isolate these two factors through an ablation study and show that the larger parent quorum accounts for most of BlueBottle's resilience to jitter, while eliminating the second voting round saves about one round of latency.

\para{Buffer rounds}
\cref{fig:eval:buffer} compares each asynchronous family across two wave lengths $w$ under network jitter. Adding a buffer round improves latency only when it recovers lost direct commits. Increasing the wave length from $w = 2$ to $w = 3$ has this effect: at $w = 2$, Nemo-Nemo Async directly commits only 78\% of its leaders (73\% with crashes), forcing the remaining slots to wait for a subsequent anchor. Restoring those direct commits at $w = 3$ offsets the extra buffer round at the median, leaving p50 latency unchanged both with and without crashes, and significantly reduces tail latency (lowering p90 by 0.5\,s without crashes and by 1.4\,s with crashes). For $w \ge 3$, however, all asynchronous configurations already directly commit at least 97\% of their leaders, so adding another buffer round (from $w = 3$ to $w = 4$ for Orcaella Async, or from $w = 4$ to $w = 5$ for Mahi-Mahi and Hybrid Mahi-Mahi) provides no further gain in direct commits and simply adds $1\rho$ to the median latency. This confirms \cref{claim:c4}.
\takeaway{4}{Under network jitter, first-round votes still form a quorum, but second-round certificates often fall short: three-round partially synchronous protocols directly commit only 7--42\% of leaders and require 10--50 rounds, whereas two-round protocols directly commit 77--81\% at the cost of roughly one extra round. Asynchronous protocols directly commit over 97\% of leaders with a single buffer round, which adds no median latency, while any further buffer round adds $1\rho$.}

\section{Related Work}
\label{sec:related}

\para{DAG-based consensus and leader scheduling}
Certified DAG protocols, such as DAG-Rider~\cite{dag-rider}, Narwhal and Tusk~\cite{narwhal}, Bullshark~\cite{bullshark}, Sailfish~\cite{sailfish}, and Shoal++~\cite{shoalpp}, disseminate each block by reliable or consistent broadcast. The resulting certificate rules out equivocation and ensures availability, at the cost of extra message delays and signatures in every round. Certified DAGs are deployed in Aptos and Aleo. Uncertified DAG protocols, which build on ideas from Hashgraph~\cite{hashgraph} and Blockmania~\cite{blockmania} and were developed in Cordial Miners~\cite{cordial} and Mysticeti~\cite{mysticeti}, instead tolerate equivocation within the DAG, and each defines a standalone commit rule and proof, whereas \sysname captures their shared decision logic in a single parameterized \decider. For leader scheduling, Carousel~\cite{carousel} introduces reputation-based leader selection for chained BFT and HammerHead~\cite{hammerhead} brings it to certified DAGs. \sysname composes scheduling with every configuration of the \decider (\cref{sec:features}).

\para{Systematizations and unified BFT frameworks}
For leader-based BFT, Bedrock~\cite{bedrock} defines a parameterized design space and execution platform for partially synchronous protocols. For DAG consensus, Raikwar et al.~\cite{sok-dag} classify protocols by consistency and availability trade-offs, but provide no common decision rule, proof, or implementation. \sysname provides a parameterized protocol framework and unified evaluation platform for uncertified DAG consensus, and pairs every configuration and feature mechanism with machine-checked proofs.

\para{Mechanized consensus verification}
Mechanized proofs of leader-based BFT establish both safety and liveness of Jolteon in Coq through LiDO~\cite{lido}. Bythos~\cite{bythos} provides compositional Coq proofs of safety and liveness for BFT protocols assembled from broadcast building blocks, whereas \sysname composes feature mechanisms around a parameterized DAG \decider (\cref{thm:compose}). For DAG consensus, Bertrand et al.~\cite{bertrand} verify safety, though not liveness, for DAG-Rider, Cordial Miners, Hashgraph, Bullshark, and an Aleph variant in TLA+ by separating DAG construction from ordering to reuse invariants. LiDO-DAG~\cite{lidodag} proves safety and liveness for Narwhal, Bullshark, and Sailfish in Rocq, and Qiu et al.~\cite{lido-mysticeti} extend LiDO-DAG to Mysticeti and identify a liveness bug caused by validators that jump rounds without building blocks. All of these DAG formalizations verify individual protocols at $n \ge 3f+1$ under Byzantine faults alone. Within Lean~4~\cite{lean4}, Veil~\cite{veil} combines SMT automation with interactive proofs to verify transition systems such as Paxos variants, reliable broadcast, and Rabia, none of which is a DAG protocol. In contrast, \sysname formalizes uncertified DAG consensus directly in Lean~4 and Mathlib~\cite{mathlib} and proves both safety and liveness for one parameterized \decider across partial synchrony and asynchrony as well as crash, Byzantine, and hybrid fault models, together with its feature compositions.

\section{Conclusion}
\label{sec:conclusion}

Many uncertified DAG protocols are in fact one protocol. We show that a single parameterized \decider recovers eight published protocols across network and fault models and yields four new ones. As a result, one proof covers every configuration, mechanism, and composition, and one implementation runs them all. Choosing a protocol reduces to trading fault tolerance against latency, at equal throughput, by changing parameters only. A blockchain therefore need not commit early to one protocol or implementation.

\para{Results not presented here}
Formalizing an entire family of protocols, rather than a single one, is a substantial effort: our Lean codebase spans about 125{,}000 lines. Work at this scale produced more results than this paper can present, and all of them are available in our formalization~\cite{leandag}. Besides the eight protocols that the \decider recovers, we have formalized Bluestreak~\cite{bluestreak} and FinWhale~\cite{finwhale}. While developing \sysname, we found attacks against Minnow~\cite{minnow} and Black Marlin~\cite{blackmarlin}. Beyond the proofs about the \decider's specification, we also prove that a concrete Lean implementation of the \decider conforms to that specification. Finally, this paper omits rate limiting, garbage collection, regenesis, and block synchronization because the formalization handles them through \emph{$k$-partial DAGs}, in which a block is valid without its full causal history. This addresses denial of service and resource control from the ground up, rather than on top of a DAG that requires a complete history.

\para{Open problem}
Our Rust implementation is not formally linked to the Lean specification or implementation. With mature theorem provers and AI agents that can produce proofs, machine-checked formalization of whole protocol families is becoming cheap, but how best to connect such proofs to Rust production code remains an ongoing challenge.

\bibliographystyle{IEEEtran}
\bibliography{references}

\appendices
\section{Security Analysis of the \decider}
\label[appendix]{app:security}

We prove that every configuration of the \decider implements atomic broadcast against its own fault model, as defined in \cref{sec:background}. Every result holds at every row of \cref{tab:params}, at every committee from the row's bound, and at every vote rule: any deterministic choice of one block of the slot from the voter's history at and above the slot's round, whenever one exists (\cref{sec:fabric:decider}). All proofs are mechanized in Lean~4, and we give the argument of each.

\begin{lemma}[Total order]
\label{lem:total-order}
If two correct validators deliver blocks $b$ and $b'$, they deliver them in the same order.
\end{lemma}
\begin{proof}[Proof sketch]
We first show that no two views decide a slot differently. Each inequality of \cref{sec:fabric:new} closes one case of comparing two decisions of the slot, by one of the counts of that section. The vote rule enters only through two facts: a block votes for at most one block of a slot, and it never blames a slot it votes in. The two Nemo-Nemo rows violate the inequality against conflicting blocks of one leader, which is needed only when a validator equivocates, and they meet its crash-only weakening. \textsc{TryDecide} outputs exactly the decisions its view derives, at consecutive slots, so two correct validators' sequences of committed leaders are prefixes of one another. Restarting the loop at the last consumed round outputs what one call on the final view would (\cref{sec:fabric:decider}). Each committed leader delivers the blocks of its causal history in a fixed order, by round and then by identifier, keeping the first block of each round and author not yet delivered, so the delivered sequences are prefixes of one another too.
\end{proof}

\begin{lemma}[Integrity]
\label{lem:integrity}
Every correct validator delivers at most one block of each round and author, and no block twice. Since blocks are signed, a delivered block of a correct validator was built by that validator.
\end{lemma}
\begin{proof}[Proof sketch]
Two blocks of a Byzantine author for one round can both lie in a committed leader's history, since blocks are not reliably broadcast. Delivery therefore keeps only the first block of each round and author, so no round and author is delivered twice, and hence no block. Every delivered block is a block of the DAG.
\end{proof}

\begin{lemma}[Agreement]
\label{lem:agreement}
If a reliable validator delivers a block $b$, every reliable validator eventually delivers $b$.
\end{lemma}
\begin{proof}[Proof sketch]
Under either network model, a reliable view is eventually contained in every other reliable view. The delivered sequence on a larger view extends the one on a smaller view (\cref{lem:total-order}), so every reliable validator eventually delivers $b$.
\end{proof}

\begin{lemma}[Validity under partial synchrony]
\label{lem:validity-ps}
After GST, every slot a correct validator leads commits on every correct view. A block of a reliable validator that every reliable validator holds at round $t$ is delivered by every reliable validator under any schedule with \emph{bounded anchor chains}, in which every slot is decided, directly or through the indirect rule, within a bounded number of rounds. Round-robin gives bounded anchor chains when $\pwavelength\,(n-|T|) < n$, $T$ the reliable set, and at Hybrid Mysticeti's committee when crashed validators are silent. When every correct validator is reliable, $n - |T| = f + c$, and one of the two conditions holds in every partially synchronous configuration of the \decider: the first everywhere except Hybrid Mysticeti with $c > 0$, and the second there.
\end{lemma}
\begin{proof}[Proof sketch]
After GST a correct leader's proposal reaches every correct validator before its timeout expires, so the correct validators build and vote on the wave, and the slot then commits on every view that holds the wave. The scan from round $t+1+\pwavelength$ reaches a committed leader, and every slot below it is decided: the slots at rounds from $t+2$ by their own scans, and those below through the indirect rule. Through weak references, every reliable block built after round $t$ reaches every block the reliable validators held at $t$, so the block is in that leader's history.
\end{proof}

\begin{lemma}[Validity under asynchrony]
\label{lem:validity-async}
At every asynchronous row, against an adversary that cannot predict the common coin, a slot followed by $M$ waves stays undecided on some correct view with probability at most $(1 - s/n)^M$, $s \ge 1$ the number of leaders a wave makes safe (\cref{sec:fabric:new}), and every block of a correct validator is delivered by every correct validator with probability 1.
\end{lemma}
\begin{proof}[Proof sketch]
A leader is \emph{safe to commit}, or \emph{safe}, if every vote-round block of its wave reaches its proposal. Every correct validator votes for a safe leader's proposal with no timing assumption, so a coin that names a safe leader commits its slot. Since every asynchronous row has $2q \ge n + f + 1$, the correct validators share a common core after two rounds, which makes $s \ge 1$ leaders safe. A wave is a fresh, independent draw of the coin, which gives the bound. A block of a correct validator eventually reaches every correct validator, whose next block references it, weakly if nothing else reaches it (\cref{sec:background}). It is therefore in the history of every later committed safe leader, and is delivered once a safe leader commits, which happens with probability 1.
\end{proof}

\para{Proof of \cref{thm:abc}}
\Cref{thm:abc} follows from \cref{lem:total-order,lem:integrity,lem:agreement,lem:validity-ps,lem:validity-async}. The lemmas prove both forms of atomic broadcast in the literature, Sailfish's Definition~1~\cite{sailfish} and Sui et al.'s Definition~2.1~\cite{sui-duan}. The forms differ mainly in two clauses: Sailfish's Integrity counts deliveries per round and author, while Sui et al.'s counts messages and requires that some replica broadcast them; and Sui et al.'s Validity requires only that the broadcaster itself deliver its message. Integrity and Total order hold in every execution, while Agreement and Validity hold for the reliable validators.
Each published row equals our formalization of its protocol, up to the vote rule (\cref{sec:fabric:existing}). \Cref{thm:abc} therefore holds for these formalizations of the published protocols.

\para{Proof of \cref{thm:compose}}
A scheduled run is a sequence of leader configurations, each in force over a range of rounds that ends at the next configuration's start. Its \emph{ledger} is the sequence of leaders the run commits in these ranges, and a validator delivers the causal histories of the ledger's leaders, through references and weak references, as in \cref{lem:total-order}.
Integrity holds of the delivered sequence of any ledger, by \cref{lem:integrity}. Validators decide every slot alike, so the next configuration, a function of committed history, is the same on every view, and two validators' ledgers up to configurations both have closed are prefixes of one another. Total order and Agreement follow as in \cref{lem:total-order,lem:agreement}.
For Validity, let configuration $k$ be in force from round $s_k$, with interval $I$, and let it be good from $s_k + \delta$, $\delta \ge 1$: its schedule has bounded anchor chains with commit gap $g$, and the reliable validators run it responsively, with crashes and restarts, past GST and the last restart. If $\delta + g \le I$, a leader of round at least $s_k + \delta$ commits before the range ends, so it is in the ledger. Configurations start at least one round apart, so $s_{t+2} \ge t+2$, and that leader's round exceeds $t+1$. Through weak references, its history holds every block the reliable validators held at round $t$.
Under asynchrony every configuration is good when the coin names a wave of safe leaders often enough, which fails with probability decaying geometrically in the number of waves, as in \cref{lem:validity-async}. Under HammerHead, the coin draws from the reputation distribution, whose support must meet every set of validators the adversary can make safe. Checkpoints finalize every slot a closed configuration commits, since they read the committed leaders.

\section{Why Two-Round Protocols Tolerate Jitter}
\label[appendix]{app:ablation}

\begin{figure}[t]
      \centering
      \includegraphics[width=\columnwidth]{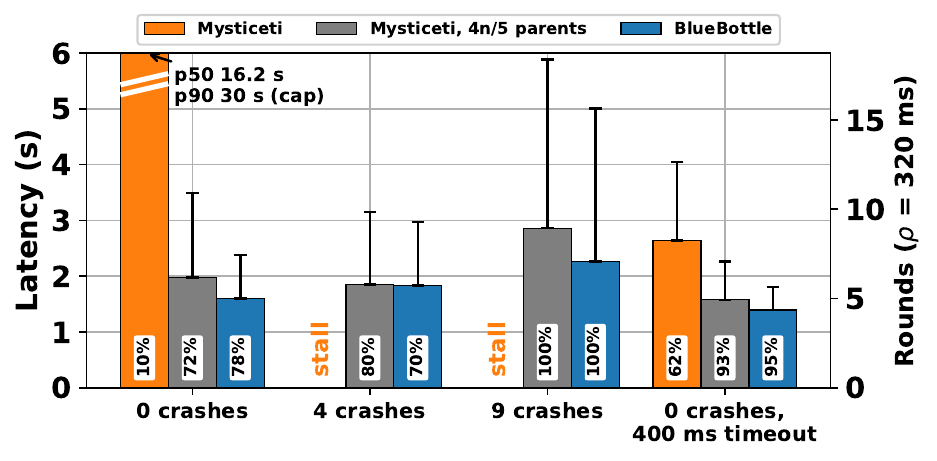}
      \caption{Parent-quorum ablation under jitter (p50, whisker to p90). Plain Mysticeti without crashes runs off the frame, and its values are printed next to it. Mysticeti with $4n/5$ parents is an ablation, not a protocol: it is not live beyond $n/5$ crashes. The number at the foot of each bar is its direct commit rate. Its p90 is capped at 30\,s. Right axis: rounds of $\rho \approx 320$\,ms, and at 9 crashes, rounds last about 620\,ms. Separate campaign from the other figures.}
      \label{fig:eval:ablation}
\end{figure}

\cref{sec:eval:jitter} leaves open why two-round $5f+1$ configurations tolerate message jitter that slows Mysticeti down to tens of rounds. Two differences between the protocols may contribute: (1) $5f+1$ configurations wait for a larger parent quorum in every round (BlueBottle advances on $4n/5 + 1 = 37$ blocks, compared to 31 for Mysticeti), and (2) they commit after a single voting round instead of requiring a second round of certificates. To separate these two effects, we evaluate an ablation that runs Mysticeti with a parent quorum of 37 blocks while keeping all of its other parameters unchanged. This configuration preserves Mysticeti's safety but loses liveness beyond $n/5 = 9$ crashes, serving purely as a diagnostic baseline. Because the ablation and BlueBottle share the same parent quorum and leader timeout, they construct the same underlying DAG: their vote counts match seed for seed with crashes, and their vote shares differ by at most 0.03 without crashes. Any latency difference between them therefore isolates the effect of the commit rule. We also re-run unmodified Mysticeti and BlueBottle in the same campaign as baselines.

\para{Effect of a larger parent quorum}
\cref{fig:eval:ablation} shows that increasing the parent quorum size accounts for most of the latency difference. Without crashes, Mysticeti's direct commit rate increases from 0.10 to 0.72 and its p50 latency drops from roughly $50\rho$ to within $1\rho$ of BlueBottle. With 4 and 9 crashes, unmodified Mysticeti stalls and commits no leaders during the 120\,s window, whereas the ablation continues to commit smoothly. (This stall does not violate Mysticeti's theoretical liveness guarantee, which holds only once the leader timeout exceeds message delays, whereas jittered delays reach 720\,ms compared to the 160\,ms timeout.) The improvement comes from the higher density of certificates rather than of votes. Specifically, the fraction of round-$(r + 1)$ blocks that vote for the leader barely changes (0.84, compared to 0.81 for unmodified Mysticeti), but each round-$(r + 2)$ block now references roughly 39 blocks from round $r + 1$ instead of 36. As a result, a typical round-$(r + 2)$ block carries about 32 votes instead of 29, clearing Mysticeti's certificate threshold $Q = 31$. The share of round-$(r + 2)$ blocks that qualify as certificates increases from 39\% to 73\%, and each slot collects an average of 34 certificates instead of 18.

\para{Effect of a single voting round}
The remaining difference between the ablation and BlueBottle reflects the second factor: committing in one voting round rather than two. At 9 crashes, the 37 remaining live validators match the parent quorum size, so every block references every live block from the previous round, and both rules directly commit every live leader and directly skip every crashed leader. In this regime, the latency difference is exactly the one-round difference in wave length, $1\rho$ (with $\rho \approx 620$\,ms). Without crashes, the difference between the two is also approximately $1\rho$.

At 4 crashes, the latency difference between the ablation and BlueBottle disappears because the two commit rules respond to crashes in opposite ways. BlueBottle requires a fixed threshold of 37 votes from a smaller set of 42 live validators, so the fraction of slots that reach 37 votes drops from 79\% (with no crashes) to 68\%, reducing its direct commit rate to 70\%. In contrast, the ablation's 37 parents cover a larger fraction of the 42 live blocks in round $r + 1$, so each round-$(r + 2)$ block carries more votes: the fraction of slots that reach the required 31 certificates rises from 69\% (with no crashes) to 80\%. The extra indirect commits incurred by BlueBottle offset the one-round advantage of its shorter wave.

\para{Effect of a longer leader timeout}
Increasing the leader timeout to 400\,ms also increases the number of votes observed by each decision-round block. While a 400\,ms timeout restores progress for unmodified Mysticeti (raising its direct commit rate to 0.62), the larger parent quorum achieves a higher direct commit rate (0.93, close to BlueBottle's 0.95) and commits roughly $3\rho$ faster.

\section{Safety Violations in Black Marlin and S-Minnow}
\label[appendix]{app:attacks}

\begin{figure}[t]
\centering
\begin{tikzpicture}[x=1.28cm,y=-0.5cm,>=latex,
  b/.style={circle,draw=gray!60,minimum size=4.2mm,inner sep=0pt,font=\tiny},
  tw/.style={b,draw=red!70!black,thick,fill=red!8},
  an/.style={b,draw=blue!60!black,thick},
  g/.style={->,green!45!black,thick,shorten >=1pt,shorten <=1pt},
  r/.style={->,red!70!black,thick,densely dotted,shorten >=1pt,shorten <=1pt},
  o/.style={->,orange!85!black,thick,dash dot,shorten >=1pt,shorten <=1pt},
  d/.style={->,blue!60!black,dashed,shorten >=1pt,shorten <=1pt},
  e/.style={->,gray!40,shorten >=1pt,shorten <=1pt}]
  \foreach \r in {0,1,2,3,4}
    \node[font=\tiny,gray] at (\r,-1.7) {round \r};
  \foreach \v in {0,1,2,3}
    \node[font=\tiny,gray,anchor=east] at (-0.4,\v) {v\v};
  \node[an] (n3) at (0,3) {3};
  \node[b] (n0) at (0,0) {0}; \node[b] (n1) at (0,1) {1}; \node[b] (n2) at (0,2) {2};
  \node[b] (n4) at (1,0) {4}; \node[b] (n5) at (1,1) {5}; \node[b] (n6) at (1,2) {6};
  \node[an] (n7) at (1,3) {7};
  \node[tw] (n8) at (2,0) {8}; \node[tw] (n12) at (2,-0.75) {12};
  \node[b] (n9) at (2,1) {9}; \node[b] (n10) at (2,2) {10}; \node[b] (n11) at (2,3) {11};
  \node[b] (n13) at (3,0) {13}; \node[an] (n14) at (3,1) {14};
  \node[b] (n15) at (3,2) {15}; \node[b] (n16) at (3,3) {16};
  \node[b] (n17) at (4,0) {17}; \node[b] (n18) at (4,1) {18};
  \node[an] (n19) at (4,2) {19}; \node[b] (n20) at (4,3) {20};
  \foreach \a/\b in {4/0,4/1,4/2,5/0,5/1,5/2,6/1,6/2,6/3,7/1,7/2,7/3,8/4,8/5,8/6,9/4,9/5,9/6,10/4,10/5,10/6,11/5,11/6,11/7,12/4,12/6,13/9,13/10,14/9,14/10,15/9,15/10,16/9,16/11,17/13,17/15,18/13,18/15,20/15,20/16}
     \draw[e] (n\a) -- (n\b);
  \draw[g] (n14) -- (n8); \draw[g] (n15) -- (n8); \draw[g] (n16) -- (n8); \draw[g] (n17) -- (n14); \draw[g] (n18) -- (n14); \draw[g] (n20) -- (n14);
  \draw[r] (n13) -- (n12);
  \draw[o] (n12) -- (n7);
  \draw[d] (n19) -- (n13); \draw[d] (n19) -- (n15); \draw[d] (n19) -- (n16);
\end{tikzpicture}
\caption{The Black Marlin execution at $n=4$, $f=1$, with validator $0$ Byzantine. Blue rings mark anchors, red rings the equivocating anchor blocks $8$ and $12$ of round $2$. Blocks $14$, $15$, $16$ reference $8$, and $17$, $18$, $20$ reference the round-3 anchor $14$ (green), so validator $X$, holding them, commits $8$ by the direct rule. Validator $Y$ lacks $17$, $18$, $20$ and fails the test. The round-4 anchor $19$ omits $14$ (blue, dashed); rounds $5$ and $6$, not drawn, commit $19$ at $Y$, whose recursion drops to round $2$ and prefers $12$, which reaches the round-1 anchor $7$ (orange), over $8$. $X$ outputs $8$, $Y$ outputs $12$, and line~27 makes each drop the other. Blocks $5$ and $7$, of correct authors, come out in opposite orders.}
\label{fig:attack-bm}
\end{figure}
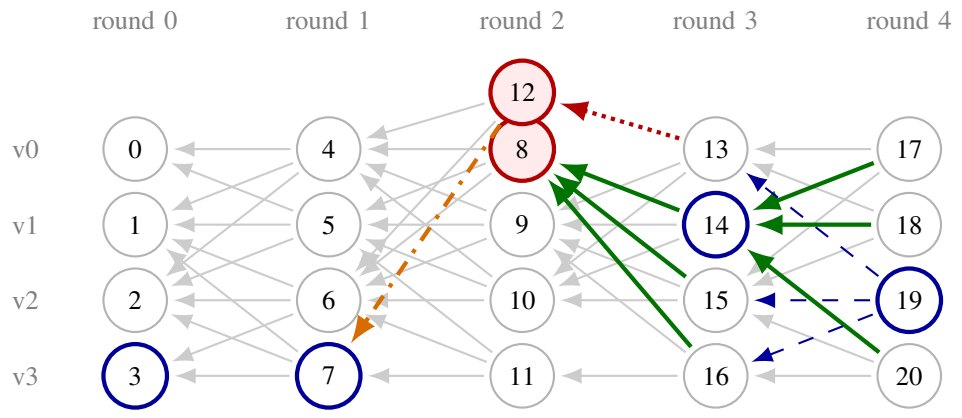

\begin{figure}[t]
\centering
\begin{tikzpicture}[x=1.9cm,y=-0.6cm,>=latex,
  b/.style={circle,draw=gray!60,minimum size=4.6mm,inner sep=0pt,font=\tiny},
  tw/.style={b,draw=red!70!black,thick,fill=red!8},
  ld/.style={b,draw=blue!60!black,thick},
  gone/.style={b,draw=gray!45,fill=gray!12},
  e/.style={->,gray!40,shorten >=1pt,shorten <=1pt},
  g/.style={->,green!45!black,thick,shorten >=1pt,shorten <=1pt},
  r/.style={->,red!70!black,thick,densely dotted,shorten >=1pt,shorten <=1pt}]
  \foreach \r in {0,1,2} \node[font=\tiny,gray] at (\r,-1.5) {round \r};
  \foreach \v in {0,1,2,3} \node[font=\tiny,gray,anchor=east] at (-0.4,\v) {v\v};
  \node[tw] (b0) at (0,0) {0}; \node[tw] (b4) at (0,-0.8) {4};
  \node[b] (b1) at (0,1) {1}; \node[b] (b2) at (0,2) {2}; \node[b] (b3) at (0,3) {3};
  \node[gone] (b5) at (1,0) {5}; \node[ld] (b6) at (1,1) {6};
  \node[b] (b7) at (1,2) {7}; \node[b] (b8) at (1,3) {8};
  \node[b] (b9) at (2,0) {9}; \node[b] (b10) at (2,1) {10};
  \node[ld] (b11) at (2,2) {11}; \node[gone] (b12) at (2,3) {12};
  \foreach \a/\b in {b5/b1,b5/b2,b7/b1,b7/b2,b8/b2,b8/b3,b6/b2,b6/b3,
                     b9/b7,b9/b8,b10/b7,b10/b8,b11/b7,b11/b8,b12/b5,b12/b7}
     \draw[e] (\a) -- (\b);
  \draw[g] (b5) -- (b0); \draw[g] (b7) -- (b0); \draw[g] (b8) -- (b0);
  \draw[r] (b6) -- (b4);
  \draw[g] (b9) -- (b6); \draw[g] (b10) -- (b6); \draw[g] (b11) -- (b6);
\end{tikzpicture}
\caption{The S-Minnow execution at $n=4$, $f=1$, with validator $0$ Byzantine and one leader per round by round-robin. Validator $0$ equivocates at round $0$ with blocks $0$ and $4$. Blocks $5$, $7$, $8$ reference $0$ (green), so $0$ carries a quorum in the whole DAG; only the round-1 leader $6$ references $4$ (red). A correct validator's view lacks $5$ and $12$ (grey): there $0$ has two supporters and is neither committed nor skipped, yet $4$ lies in the history of $6$, which $9$, $10$, $11$ reference, so $6$ commits. When $5$ arrives, $0$ passes Quorum, and its slot precedes the slot of $6$ already output.}
\label{fig:attack-mn}
\end{figure}
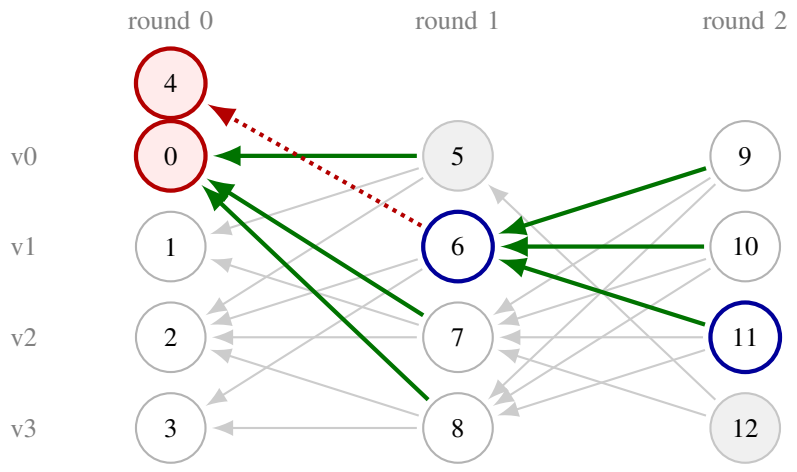

Black Marlin~\cite{blackmarlin} and S-Minnow~\cite{minnow} are partially synchronous uncertified DAG protocols that commit from less of the DAG than Mysticeti at $n \ge 3f+1$. Both decide a leader slot by a direct rule, which commits a block that a quorum of the round above references, and an indirect rule, which decides earlier slots once a later leader commits and tests only whether a block of the slot lies in that leader's history. When the leader equivocates, the two rules can choose different blocks. \Cref{fig:attack-bm,fig:attack-mn} give an execution of each at $n = 4$, $f = 1$.

\para{Black Marlin} elects one anchor per round. Its direct rule (line~16 of Algorithm~1~\cite{blackmarlin}) commits the anchor of round $r-2$ when it has $n-f$ supporters and a supported anchor of round $r-1$ references it; a validator that fails the test never retries the round. The commit recursion then delivers the anchor's history, recursing on one anchor per round chosen by a metric that ignores support, and skips any block whose author and round it already delivered.

\para{S-Minnow} commits a leader block $l$ when $2f+1$ validators reference it from the round above, and every earlier slot holds a block that lies in $l$'s history, or is concurrent with $l$ and committed or skipped (Definition~9 of~\cite{minnow}). The first case clears a slot without deciding it.

\end{document}